\documentclass[10pt]{article}
\usepackage{amsmath}
\usepackage{amssymb}
\usepackage{graphicx}
\usepackage{cite}
\usepackage{color}
\usepackage{setspace}
\usepackage[labelfont=bf,labelsep=period,justification=raggedright]{caption}
\usepackage{psfrag}
\usepackage{amsthm}

\makeatletter
\renewcommand{\@biblabel}[1]{\quad#1.}
\makeatother

\date{}
\newcommand{\xx}{{\boldsymbol{x}}}
\newcommand{\FF}{{\boldsymbol{F}}}
\newcommand{\xixi}{{\boldsymbol{\xi}}}

\usepackage{soul}

\begin{document}
\begin{flushleft}
{\Large
\textbf{$3$D Hybrid Modelling of Vascular Network Formation}}
\\
Holger Perfahl$^{1,\ast}$, Barry D. Hughes$^2$, Tom\'as Alarc\'on$^{3,4,5}$, Philip K. Maini$^6$, Mark C. Lloyd$^7$, Matthias Reuss$^1$, Helen M. Byrne$^6$
\\
\bf{1} Center Systems Biology, University of Stuttgart, Stuttgart, Germany
\\
\bf{2} School of Mathematics and Statistics, University of Melbourne, Australia
\\
\bf{3} ICREA, Pg. Llu\'{\i}s Companys 23, 08010 Barcelona, Spain
\\
\bf{4} Centre de Recerca Matem\`atica, Campus de Bellaterra, Barcelona, Spain
\\
\bf{5} Departament de Matem\`atiques, Universitat Aut\`onoma de Barcelona, Barcelona, Spain
\\
\bf{6} Wolfson Centre for Mathematical Biology, Mathematical Institute, University of Oxford, Oxford, UK
\\
\bf{7} H. Lee Moffitt Cancer Center and Research Institute, Tampa, USA
\\
$\ast$ E-mail: holger.perfahl@ibvt.uni-stuttgart.de\\
\end{flushleft}
\begin{abstract}
We develop an off-lattice, agent-based model to describe vasculogenesis, the {\it de novo} formation of blood vessels from endothelial progenitor cells during development. 
The endothelial cells that comprise our vessel network are viewed as linearly elastic spheres that move in response to the forces they experience. We distinguish two types of endothelial cells: {\it vessel elements} are contained within the network and {\it tip cells} are located at the ends of vessels. 
Tip cells move in response to mechanical forces caused by interactions with neighbouring vessel elements and the local tissue environment, chemotactic forces and a persistence force which accounts for their tendency to continue moving in the same direction.
Vessel elements are subject to similar mechanical forces but are insensitive to chemotaxis. 
An angular persistence force representing interactions with the local tissue is introduced to stabilise buckling instabilities caused by cell proliferation.
Only vessel elements proliferate, at rates which depend on their degree of stretch: elongated elements have increased rates of proliferation, and compressed elements have reduced rates. Following division, the fate of the new cell depends on the local mechanical environment: the probability of forming a new sprout is increased if the parent vessel is highly compressed and the probability of being incorporated into the parent vessel increased if the parent is stretched.

Simulation results reveal that our hybrid model can reproduce the key qualitative features of vasculogenesis.   
Extensive parameter sensitivity analyses show that significant changes in network size and morphology are induced by varying the chemotactic sensitivity of tip cells, and the sensitivities of the proliferation rate and the sprouting probability to mechanical stretch. Varying the chemotactic sensitivity directly influences the directionality of the networks. The degree of branching and thereby the density of the networks is influenced by the sprouting probability.   
Glyphs that simultaneously depict several network properties are introduced to show how these and other network quantities change over time and also as model parameters vary. 
We also show how equivalent glyphs constructed from {\it in vivo} data could be used to discriminate between normal and tumour vasculature and, in the longer term, for model validation.
We conclude that our biomechanical hybrid model can generate vascular networks that are qualitatively similar to those generated from {\it in vitro}  and {\it in vivo} experiments.  
\end{abstract}
\section{Introduction}
\label{Intro}
In order to ensure their continued survival, many biological tissues are endowed with a network of blood vessels that delivers vital nutrients such as oxygen and glucose to cells, removes waste products and facilitates information exchange between different organs \cite{folkman1995angiogenesis}. The vessel networks typically form via angiogenesis and/or vasculogenesis \cite{risau1997mechanisms}. 
The process of angiogenesis has been widely studied due to its importance in wound healing and tumour growth, angiogenesis marking the transition from the relatively harmless and localised phase of avascular tumour growth to the potentially life-threatening phase of vascular growth \cite{carmeliet2005angiogenesis}. During angiogenesis, new blood vessels emerge from pre-existing, perfused vessels, the endothelial cells that constitute the vessels being stimulated to proliferate and migrate chemotactically in response to growth factors produced by cells that lack an adequate supply of nutrients.
In contrast, vasculogenesis is the {\it de novo} formation of new blood vessels from isolated endothelial cells. As such, it is not reliant upon the presence of a pre-existing vascular network and is a prominent feature of embryonic development.  
While in practice, both vasculogenesis and angiogenesis may simultaneously participate in vascular network formation during wound healing, tumour growth and embryonic development, their relative contributions remain keenly debated
\cite{drake2003embryonic}. As a result, increased understanding of both processes and their interactions is urgently needed. Such understanding may also enable experimentalists and clinicians to establish how best to combine vascular targeting agents with other treatments either to stimulate healing of chronic wounds or to arrest the growth of solid tumours. 

There is an extensive theoretical literature devoted to mathematical and computational modelling of vascular network formation. Models have been developed across the spectrum of physiological space and time scales, using a variety of frameworks. 
The most widely used methods are based on
ordinary differential equations, partial differential equations and/or agent-based approaches. They differ in the geometrical resolution and detail they represent. Ordinary differential equation models describe the time evolution of global quantities, such as the number of vessels and the tumour volume \cite{hahnfeldt1999tumor}. Models formulated using partial differential equations typically 
describe the time evolution of spatially distributed, macroscale features, such as vessel volume fractions and the concentrations of oxygen and chemoattractants \cite{Hubbard_2013}, although more recent, phase-field models provide a framework for simulating morphological features of vascular networks with continuous variables \cite{vilanova2013phasefield}. Agent-based approaches permit a more detailed study of the biological phenomena, on a scale at which the spatial and temporal evolution of individual blood vessels and/or endothelial cells may be resolved. 
In the paragraphs that follow, we illustrate briefly each of the aforementioned categories, focussing on agent-based approaches, which are the subject of the present work. 

Hahnfeldt et al. \cite{hahnfeldt1999tumor} \textcolor{black}{and Arakelyan et al. \cite{arakelyan2003}} proposed two-compartment ordinary differential equation models for the growth of a vascular tumour and its response to treatment with an anti-angiogenic chemical. More generally, spatially-resolved, continuum models feature partial differential equations (PDEs) 
for the endothelial cell volume fraction that are coupled to PDEs for chemoattractants, chemorepellents and the extracellular matrix. The key advantages
of such continuum models are the relatively small number of parameters that they contain and that they can be simulated efficiently. The drawbacks are that individual cell properties, geometrical details of the vascular morphology and subcellular networks cannot be included easily. Following a continuum approach, Balding and McElwain \cite{balding1985mathematical} developed a model including densities of blood vessels and capillary tips, that describes sprout formation and fusion from a pre-existing network in response to tumour angiogenic factors. This model applies the ``snail-trail'' concept under which moving capillary tips leave behind blood vessels. Flegg et al. developed a similar,  continuum model with three-species (oxygen concentration, capillary tip density, blood vessel density) to study the efficacy of hyperbaric oxygen therapy for
healing chronic wounds \textcolor{black}{\cite{flegg2009three,machado2011}}. \textcolor{black}{Several modelling papers have investigated the effects that deformation of the extracellular matrix have on network structure. For example, Edgar et al. \cite{edgar2013} have shown how fibre orientation may guide the movement of tip cells. Other approaches based on mechano-chemical models \cite{manoussaki1996}, continuum \cite{namy2004,tosin2006}, and continuum-discrete modelling \cite{stephanou2015}, have tackled this issue. For example, Dyson et al. have used a multiphase approach to show how fibres embedded in the tissue matrix may bias cell movement and how cell movement may deform the fibres \cite{Dyson2015}. However,  
none of these models couples the mechanical stress that the endothelial cells experience to their growth, proliferation and the phenotype of their daughter cells.} As an example of an agent-based off-lattice angiogenesis model we refer to  \textcolor{black}{Stokes and Lauffenberger} \cite{stokes1991analysis} which includes sprout movement as a biased persistent random walk, based on a stochastic differential equation for the cell acceleration. \textcolor{black}{See also Anderson and Chaplain \cite{anderson1998} and Plank and Sleeman \cite{plank2004}}. The persistence of motion is directly linked to chemotaxis.
More information about these approaches can be found in the review articles of Mantzaris et al.  \cite{mantzaris2004mathematical}, Ambrosi et al. \cite{ambrosi2005review}, Merks and Koolwijk \cite{merks2009modeling}, and Scianna et al. \cite{scianna2012review}. 
 
We now highlight some existing agent-based models of vasculogenesis.  Bentley~et~al. \textcolor{black}{\cite{bentley2009tipping}} resolve the shape and movement of vessel sprouts: each cell is decomposed into a number of connected agents and attention focusses on the influence of delta-notch-signalling on the initiation of vascular sprouts. Their work focusses on small segments of vascular networks. Other authors have simulated vasculogenesis and vascular network formation on larger length scales. For example, Merks et al.  \cite{merks2008contact} have used the Cellular Potts Model (CPM) to investigate how contact inhibition influences the movement of capillary sprouts in response to an extracellular, diffusible chemoattractant produced by the endothelial cells. Cell movement, shape and alignment are determined by minimising an energy function that accounts for cell-cell binding, volume constraints, and chemotaxis, with chemotactic movement restricted to lattice sites that are adjacent to sites occupied by extracellular 
matrix (ECM). Scianna~et~al.~\cite{scianna2011multiscale} also use the CPM to develop a multiscale model that accounts for cell activation, migration, polarisation and adhesion, and in which each cell is decomposed into a nuclear and a cytosolic region. In the intracellular space the concentrations of arachidonic acid, nitric oxide and calcium are described by reaction-diffusion equations. The energy functional that determines cell morphology and movement accounts for the shape (area, perimeter), adhesion energy (intraellular, and nuclei and cytosol within the same cell), chemotaxis, 
contact inhibition, and persistence in cell movement. The extracellular concentration of vascular endothelial growth factor (VEGF) is described by a reaction diffusion equation. Building on the CPM, Szabo~et~al.~\cite{szabo2007network,szabo2008multicellular,szabo2012invasion} studied the roles of cell elongation and cell-matrix interactions on vascular network formation. In other work, Oers and Merks~\cite{van2013mechanical} coupled a CPM to a finite element model for substrate deformation. The energy function in the CPM depended on the strain and orientation-dependent stiffness of the extracellular matrix. Their model simulations yielded realistic patterns of network formation and sprouting from clusters of endothelial cells.

In contrast to the agent-based models mentioned above, where a single cell is represented by several agents, cellular automata represent one cell by one agent. For example, St\'ephanou \cite{stephanou2005mathematical} developed an agent-based framework in which the movement of individual cells is influenced by chemotaxis and haptotaxis. In their model, interactions between vessels and the extracellular matrix play an important role in regulating cell movement. 
In addition to changes in the structure of the network, vessel radii also adapt in response to hydrodynamic, metabolic and angiogenic stimuli. Watson et al. adapted this approach to develop a hybrid two-dimensional model of retinal vascular plexus development \cite{watson2012dynamics}. The model accounts for individual astrocytes and endothelial tip cells that move on a lattice, and uses a continuum approach to model the distribution of biochemical signalling molecules and ECM components. Building on earlier work by Alarc\`on et al. \cite{alarcon2005, owen2009}, Perfahl et al. \cite{perfahl2011multiscale}, \textcolor{black}{Macklin et al. \cite{macklin2009}, Shirinifard et al. \cite{shirinifard2009}, and Welther et al. \cite{welter2008,welter2009}} developed a similar cellular automaton model of vascular tumour growth. 

A common feature of the cellular automaton model mentioned above is that cell-cell interactions are rule-based: mechanical forces are not included. Additionally,  
network nodes are restricted to lattice sites, creating angular, non-smooth networks. Off-lattice models can be used to address these limitations and also generalise naturally to three dimensions. They have been widely used to investigate cell-cell interactions in multicellular systems, including the liver \cite{hoehme2010prediction}, development \cite{mclennan2012multiscale}, avascular tumour spheroids \cite{schaller2005multicellular} and homeostasis in the intestinal crypt \cite{pitt2009chaste}. 
A mechanically-based, hybrid model for network formation was proposed by Jackson and Zheng~\cite{jackson2010cell}. In their approach, tip and stalk cells are modelled as interconnected, elastic agents. Chemotactic movement of the tip cells produces mechanical forces that act on the neighbouring stalk cells whose rates of proliferation depend on their mass and maturity. An alternative off-lattice agent-based model of angiogenesis and vascular tumour growth was presented in Drasdo~et~al.~\cite{Drasdo2009}.

\begin{figure}[ht]
\begin{center}
\includegraphics[width=13.5cm]{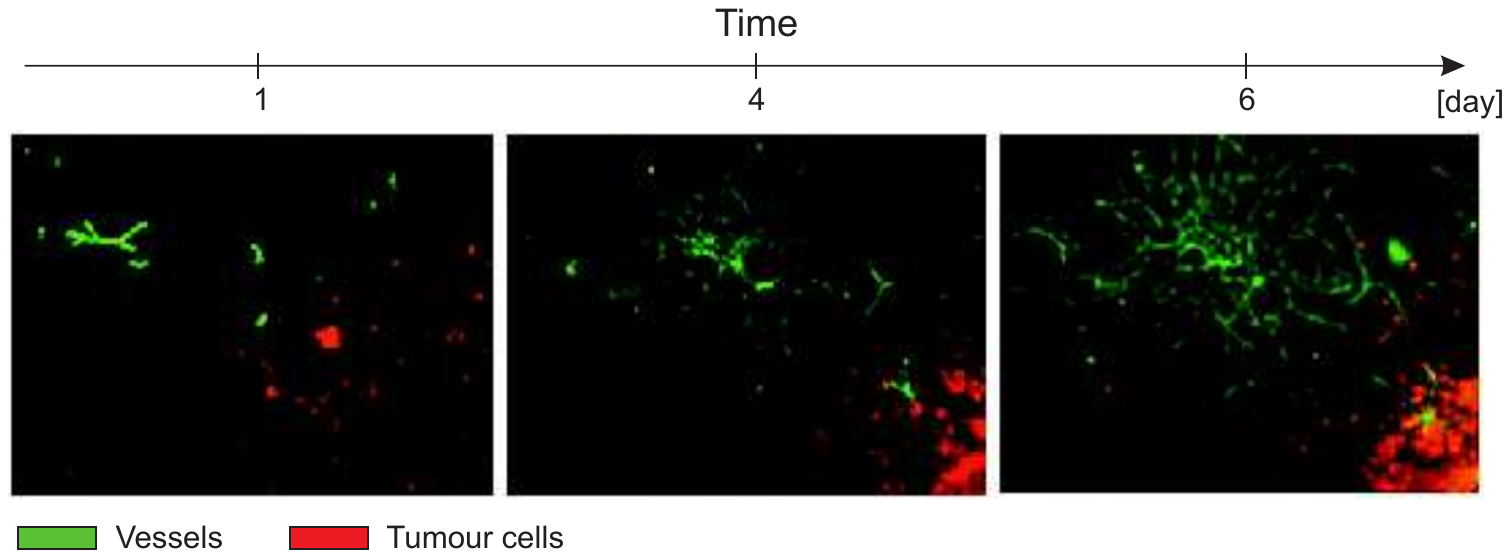}\\
\caption{{\bf Series of images showing how vascular networks emerge from rat microvessel segments {\it in vitro}}. Isolated green fluorescent rat microvessel segments were embedded within matrigel that sits on top of a droplet containing red fluorescent breast cancer cells. The endothelial cells proliferate and migrate in response to tumour-derived growth factors, forming small networks.
{\color{black} The images were collected on days 1, 4 and 6 (left-to-right).} 
The green and red scale bars correspond to length scales of 500$\mu$m. }
\label{fig:MoffittImage}
\end{center}
\end{figure}

{\color{black} Similay to Drasdo~et~al.~\cite{Drasdo2009}, and in contrast to the two-dimensional, lattice-based models described above,} 
in this paper we develop a three-dimensional (3D), off-lattice, hybrid model of vasculogenesis, although the cells (or agents) are assumed to have fixed shape, for simplicity and to improve the computational efficiency. As such, blood flow is not considered as
part of our model. Further, we do not describe the individual endothelial cells that line a small diameter blood vessel; we use spherical elements to
represent small vessel sections. 
Our model is motivated by {\it in vitro} experiments in which isolated green fluorescent rat microvessel segments were embedded within matrigel that sits on top of a droplet containing immortalised red fluorescent MDA-231 breast cancer cells so that the tumour and endothelial cells were not in direct contact. 
{\it In vitro} images, acquired on days 1, 4 and 6 using a Zeiss Z1 Observer microscope at 5x magnification, reveal endothelial cell proliferation and migration in response to tumour-derived growth factors, and the formation of small unperfused networks (see Figure \ref{fig:MoffittImage}). 
Further motivation for our hybrid model comes from experiments designed to understand how endothelial cells respond to mechanical stretch
\cite{Liu:2007,Zheng:2008}. In \cite{Liu:2007}, Liu and colleagues show that endothelial cells increase their rate of proliferation under stretch and that 
both cell-cell adhesion and engagement of vascular endothelial cadherin are needed to transduce the mechanical stretch into proliferative signals. 
In separate work \cite{Zheng:2008}, Zheng et al. showed that endothelial cells response to static stretch by increasing their rates of cell
proliferation, lumen formation and branching, and that VEGF binding is necessary to mediate these responses.

Our agent-based, off-lattice model represents the vasculature as a network of spheres whose centres are connected by linear springs and whose movement 
in 3D is determined by applying local force balance. We distinguish two types of endothelial cell agents: {\it vessel elements} which are contained within the network and proliferate, and {\it tip cells} which are located at the ends of vessels, do not proliferate and respond via chemotaxis to spatial gradients in angiogenic factors such as VEGF. 
For simplicity, we assume that tip cells and vessel elements neither change phenotype nor swap position, although there is experimental evidence for these
phenomena \cite{bentley2009tipping}.
In contrast to existing hybrid models and motivated by \textcolor{black}{work of Liu et al. \cite{Liu:2007} and Zheng et al. \cite{Zheng:2008}}, here cell proliferation and branching are assumed to depend on the degree of mechanical stretch (or compression) experienced by individual vessel elements. 
Thus, two key processes drive endothelial cell movement: chemotactic movement of tip cells creates a {\it pull} which acts on vessel elements contained within the network whereas cell proliferation creates a mitotic {\it push} which acts on the tip cells.

Numerical simulations reveal that our hybrid, mechano-chemical model can reproduce the key features of vasculogenesis. Extensive computations are performed in order to show how parameter changes influence network development. Since model simulations are stochastic, exact
network comparisons are not possible. Instead,
our parameter sensitivity analyses are based on network features extracted from multiple simulations generated with different random seeds. Metrics used to characterise the networks include the following: the total network length, the average number of branch points per unit length, the tortuosity and the distribution of vessel segment lengths. We note that these metrics 
can also be extracted from {\it in vitro} and {\it in vivo} experimental data and could, thereby, facilitate comparisons between our model and available data. Since variation of model parameters can affect multiple metrics, we also introduce glyphs simultaneously to visualise several network metrics. 
{\color{black} A glyph is a graphical object whose attributes are bound to 
data \cite{maguire2012taxonomy}. The two-dimensional glyphs that we develop enable us to clearly present multidimensional data or metrics in a single graphical entity.}

The main achievements of the paper can be summarised as follows:
\begin{itemize}
\item The development of an off-lattice hybrid model of vasculogenesis in which mechanical stretch regulates endothelial cell proliferation and capillary sprout formation;
\item The identification of quantitative metrics that can be used robustly to characterise and compare vessel networks and to study the impact of external perturbations on network structure;
\item The design and use of glyphs as an objective way of aggregating multiple network features in one diagram.
\item A demonstration that mechanical stimuli alone can generate networks whose morphological features are qualitatively 
similar to those observed {\it in vitro} and {\it in vivo};
\end{itemize}

The remainder of the paper is structured as follows. The mathematical model is introduced in Section~\ref{sec:methods}. Simulation results are presented in Section~\ref{sec:results}. The paper concludes in Section~\ref{sec:discussion} where we discuss our findings and outline directions for future work.


\section{Methods}\label{sec:methods}
In this section we introduce the 3D agent-based model that we have developed, using an off-lattice approach, to simulate vasculogenesis. Our goal is to establish whether physically realistic vessel networks can be generated when endothelial cell proliferation and capillary sprout formation are regulated by mechanical effects. 
\textcolor{black}{A detailed description of the computational model is included below while information about the algorithm and parameter values used to generate numerical simulations is provided in the Supplementary Materials}. 
In order to characterise objectively morphological changes that occur as the networks evolve and/or system parameters are varied,
several metrics are introduced (e.g. total vessel length, tortuosity and number of branch points per unit vessel length) and applied to the networks. Glyphs that simultaneously depict multiple metrics are also introduced and used to present, in a concise manner, network attributes.

{\color{black} 
\subsection{The Computational Model}\label{sec:model}
\begin{figure}[ht]
\begin{center}
\includegraphics[width=11.5cm]{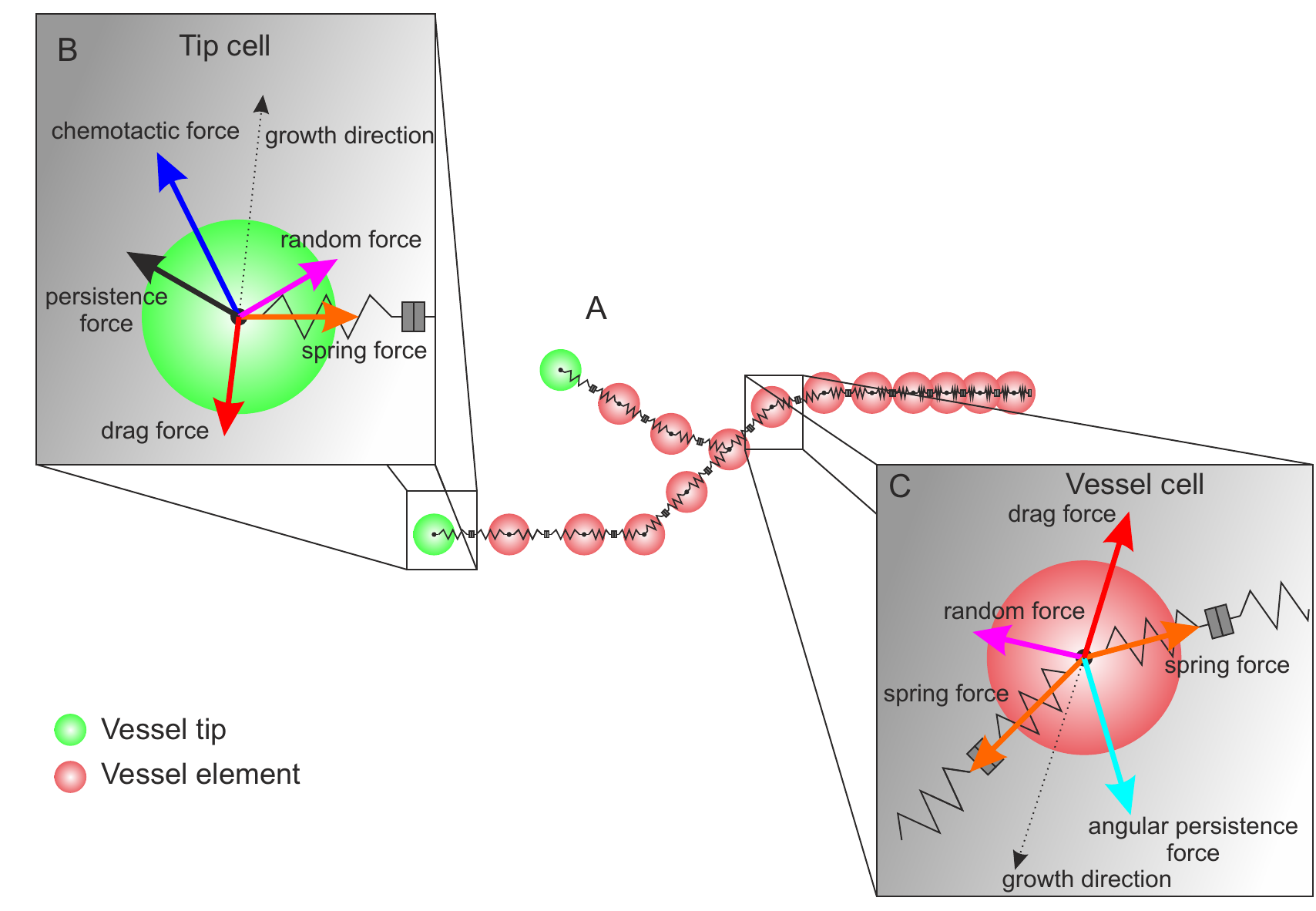}\\
\caption{{\bf Schematic of a typical vessel segment, highlighting the forces that act on its constituent elements.} {\bf (A)} A vessel segment consists of a series of inter-connected vessel elements (red), with tip cells (green) at their ends. {\bf (B)} Tip cells are subject to spring forces due to cell-cell interactions, random forces due to cell-tissue interactions, chemotactic forces due to gradients in angiogenic factors, directional persistence forces and drag. {\bf (C)} Vessel elements are subject to spring forces, random forces, an angular persistence force and drag. }
\label{fig:VesselModel1}
\end{center}
\end{figure}
Our three-dimensional, off-lattice, agent-based model aims to describe the {\it de novo} formation of vascular networks that occurs during vasculogenesis. We distinguish two types of endothelial cells: {\it tip cells} and {\it vessel elements} (see Figure~\ref{fig:VesselModel1}). 
Tip cells are located at the blunt end of each capillary sprout and all other (internal) segments are vessel elements. 
While recent experimental results have shown that tip cells may change position (and phenotype) with endothelial cells located in the same sprout \cite{hellstrom2007dll}, here, for simplicity, we neglect such effects and assume that the identity of the leading tip cell in a particular sprout is fixed.
Tip cells and vessel elements are modelled as linearly elastic spheres and all pairs of connected cells or elements exert equal and opposite (mechanical) forces on each other.

In our model, tip cells and vessel elements differ in two important ways (see Figure~\ref{fig:VesselModel1}): vessel elements can proliferate whereas tip cells cannot; tip cells are subject to a chemotactic force caused by spatial gradients in local levels of the diffusible angiogenic factor, VEGF, whereas vessel elements are not sensitive to VEGF (cell-cell contacts have been shown to inhibit VEGF-induced signalling within a vessel \cite{merks2008contact}). 
Thus, as shown in Figure~\ref{fig:VesselModel1}(A), tip cells perform a persistent random walk, biased by chemotaxis in response to spatial gradients in VEGF and constrained by mechanical forces due to (linearly elastic) cell-cell interactions and drag forces due to interactions with the local tissue matrix. Vessel elements are subject to drag, mechanical and random forces and an angular persistence force, the latter two forces accounting for cell interactions with the local environment. We introduce below the functional forms used to model each force but first explain how we derive the equations of motion for each vessel element.

\begin{table}[hptb]
\caption{{\bf Model assumptions.} 
Summary of the key differences between tip cells and vessel elements. 
Key: $+$ and $0$ indicate whether a cell type exhibits a particular property ($+$) or not ($0$).}
\centering
\label{tab:assumptions} 
\begin{tabular}[t]{lcc}
\hline\noalign{\smallskip}
{\bf Property} & {\bf Tip cell} & {\bf Vessel element} \\[3pt]
Chemotaxis &+ &0\\
Angular persistence &0 &+\\
Directional persistence of movement &+ &0\\
Stress-dependent proliferation &0 &+\\
\noalign{\smallskip}\hline
\end{tabular}
\end{table}
%
%

Suppose that at time $t$, the network comprises $N = N(t)$ elements (tip cells and vessel elements) that are located within a three-dimensional, Cartesian domain of size $W_X \times W_Y \times W_Z$. 
We denote by $\xx_i(t)$ the position of the centre of vessel segment $i \: (i=1, \dots, N)$, and record in 
an adjacency matrix $\mathcal{E}$ the node numbers of all pairs of connected vessel elements.
We use Newton's second law to derive the equations of motion.
In the over-damped limit, we neglect inertial effects and obtain the following force balances for tip cell $i$ and vessel segment $j$ respectively:
\begin{equation}\label{tip:movement}
\mu \frac{\mathrm{d} \xx_i}{\mathrm{d}t}=\FF_i^m+\FF_i^r+\FF_i^c+\FF_i^p,
\end{equation}
\begin{equation}\label{vessel:movement}
\mu \frac{\mathrm{d} \xx_j}{\mathrm{d}t}=\FF_j^m+\FF_j^r+\FF_j^a.
\end{equation}
In equations (\ref{tip:movement}) and (\ref{vessel:movement}), we assume that the drag force on vessel element $i$ is proportional to its velocity $\mathrm{d} \xx_i / \mathrm{d}t$, the positive constant $\mu$ denoting the drag coefficient. We denote the mechanical force by $\FF_i^m$, the random force by $\FF_i^r$, the chemotactic force by $\FF_i^c$, and the directional and angular persistence forces by $\FF_i^p$ and $\FF_i^a$ respectively. These forces are prescribed as follows.

\begin{itemize}
\item {\bf Mechanical force, $\FF_i^m$} {\it(tip cells and vessel elements)}\\
The mechanical force acting on a tip cell or vessel element is the net force exerted on it by its neighbours. We assume that cells/elements $i$ and $j$ only interact if the distance between their centres is less than a fixed value, $l_c$. In more detail, and following the interacting sphere approach outlined in 
\cite{drasdo2005single,drasdo2001individual,meineke2001cell,walker2004epitheliome},
if $|\xx_i-\xx_j|<l_c$, then the interaction force between cells/elements $i$ and $j$ is parallel to the vector $\xx_i-\xx_j$ connecting their centres and its magnitude $S$ is piecewise linear, taking values that range between  $S=S_c$ for highly compressed cells/elements, and $S = 100 S_c$ for highly stretched cells/elements. 
 
If we denote by $N_i$ those cells/elements $j \neq i$ for which $(i,j) \in \mathcal{E}$ and $|\xx_i-\xx_j|<l_c$, then the net mechanical force acting on cell/element $i$ is given by:
\begin{equation}\label{mechanicalForce}
\FF_i^m=\sum_{\substack{j\in N\\j\neq i}}S(2R_c-|\xx_i-\xx_j|) \: 
\frac{(\xx_i-\xx_j)}{|\xx_i-\xx_j|}.
\end{equation}
In equation (\ref{mechanicalForce}), $R_c$ is the target cell/element radius and the spring function $S(x)$
is a piecewise linear function 
\[ S(x) = \begin{cases}
S_c x &\mbox{if} \;\; x > 0, \\
100S_c x &\mbox{if} \;\; 2R_c-l_c< x \leq 0,\\
0&\mbox{else}, \end{cases} \]
so that the interaction force depends on the distance between the cells/elements and is much stronger, 
and attractive, for pairs of stretched cells/elements ($|\xx_i-\xx_j|>2R_c$) than it is for pairs of compressed cells/elements. We note that $l_c>2R_c$ holds for the cut-off distance $l_c$.
\item {\bf Random force, $\FF_i^r$} {\it(tip cells and vessel elements)}\\
We assume that all vessel elements experience random forces due to heterogeneity in the surrounding tissue environment and that the random force acting on vessel element $i$ is given by:
\begin{equation}\label{randomForce}
\FF_i^r=\sigma \xixi_i,
\end{equation}
where $\xixi_i$ is a unit vector ($|\xixi_i|=1$), pointing in a direction chosen randomly from a uniform distribution, and the constant $\sigma$ denotes the vessel element's sensitivity to random fluctuations.
\item {\bf Chemotactic force, $\FF_i^c$} {\it(tip cells)}\\
Following \cite{merks2008contact,scianna2011multiscale}, we suppose that cell-cell contact inhibits the chemotactic sensitivity of vessel elements so that only tip cells are chemotactic.
We assume further that the chemotactic force acting on tip cell $i$ is proportional to the gradient of the VEGF concentration $c=c(t,\xx_i)$ at its centre so that
\begin{equation}\label{chemotacticForce}
\FF_i^c=\chi \nabla c(t,\xx_i),
\end{equation}
where the positive constant $\chi$ denotes the chemotactic sensitivity. 
For simplicity, in the simulations that follow we prescribe a fixed chemoattractant gradient, with $\nabla c=(-c_x,0,0)$ so that the constant $c_x$ denotes the VEGF gradient.
\item {\bf Directional persistence force, $\FF_i^p$} {\it(tip cells)}\\
We suppose that tip cells have a tendency to continue moving in the same direction.
Rather than retaining inertial effects in Equations (\ref{tip:movement}) and (\ref{vessel:movement}), we model this tendency by introducing the directional persistence force.
Unlike inertia, we assume that this force is induced by active cell movement 
(along the ECM, for example) and that it acts in the direction of cell movement over a timescale of 
length $\tau > 0$. Accordingly, we denote the directional persistence force, $\FF_i^c$ acting on tip
cell $i$ as follows:
\begin{equation}\label{persistenceForce}
\FF_i^p= \omega_p \: \frac{\xx_i(t)-\xx_i(t-\tau)}{|\xx_i(t)-\xx_i(t-\tau)|},
\end{equation}
where $\omega_p > 0$ denotes the persistence coefficient. We remark that
a similar approach was adopted in \cite{stokes1991analysis} where persistence was directly linked with chemotaxis.
\item{\bf Angular persistence force, $\FF^a_i$} {\it(vessel elements)}\\
The angular persistence force accounts for forces that vessel elements experience due to interactions with their microenvironment. We assume that it acts to stabilise buckling instabilities induced by proliferation (see Figure \ref{fig:AngularPersistence2}). 
If vessel element $i$ is not a branching point, and has two neighbours $j$ and $k$ (so that $(i,j),(i,k)\in\mathcal{E}$), then the angular persistence force acting on vessel element $i$ is given by
\begin{equation}\label{angularpersistenceForce}
\FF^a_i=\omega_a(\pi-\alpha_{\mathrm{angular}})\frac{(\xx_j-\xx_i)+(\xx_k-\xx_i)}{|(\xx_j-\xx_i)+(\xx_k-\xx_i)|}.
\end{equation}
In equation (\ref{angularpersistenceForce}), the positive constant $\omega_a$ denotes the angular spring constant, and $0 \leq \alpha_{\mathrm{angular}} \leq \pi$ is the (smallest) angle between the vectors $(\xx_j-\xx_i)$ and $(\xx_k-\xx_i)$. 
If vessel element $i$ is a branch point, connected to more then two other elements, then $\alpha_{\mathrm{angular}}$ is taken to be
the smallest branching angle between the vessel elements. 
\begin{figure}[ht]
\begin{center}
\includegraphics[width=11.5cm]{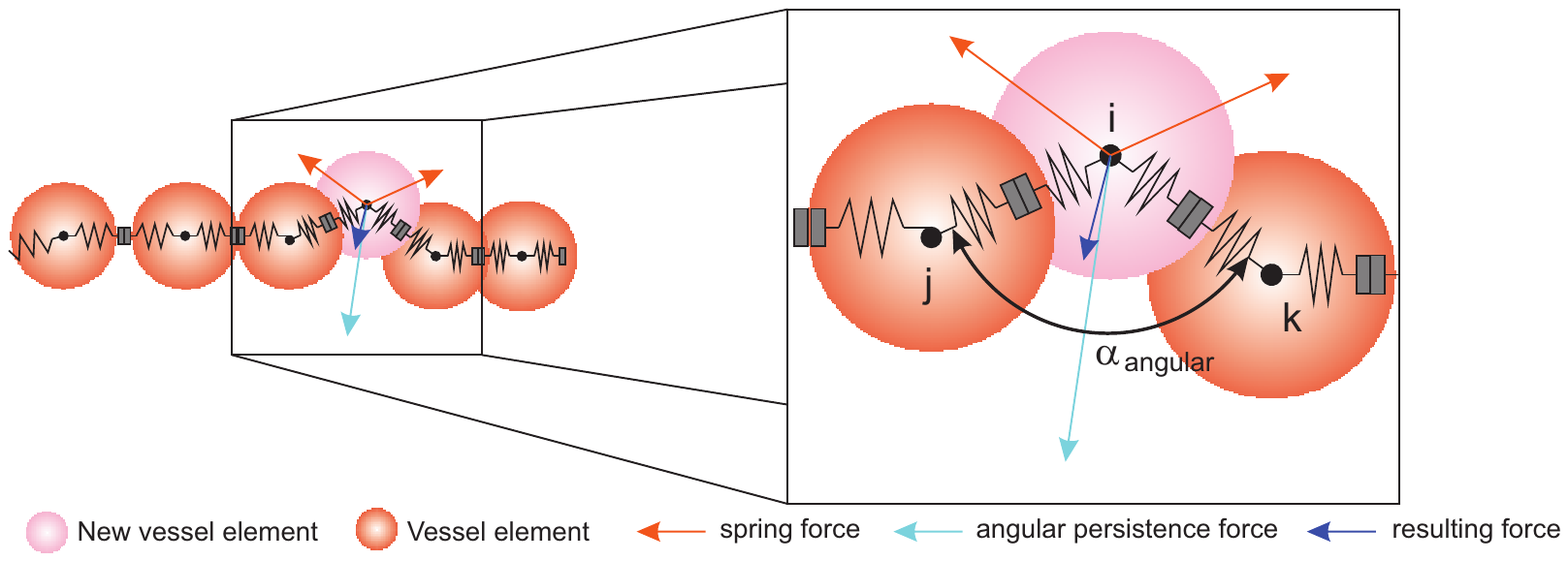}\\
\caption{{\bf Schematic of the angular persistence force.}  Suppose vessel element $i$ is connected to vessel elements $j$ and $k$ and that their centres have coordinates $\xx_i$, $\xx_j$, and $\xx_k$, respectively. Then the angle $\alpha_{\mathrm{angular}}$ is defined to be the smallest angle between the two vessel segments, as indicated. This angle and the coordinates of the three vessel elements are used in Equation (\ref{angularpersistenceForce}) to determine the angular persistence force, $\FF^a$.
}
\label{fig:AngularPersistence2}
\end{center}
\end{figure}
\end{itemize}

We complete our model of vasculogenesis by explaining how division and sprouting are represented. We assume that both processes are regulated by the amount of stretch that a vessel element experiences. Thus, for example, an elongated vessel element increases its rate of progress through the cell cycle while a compressed element decreases its rate. For simplicity, and extending an approach used in Owen~et~al.~\cite{owen2011mathematical}, the progress of vessel segment $i$ through the cell cycle is monitored by a phase variable $\phi_i(t) \in [0,1]$ whose evolution is determined by the following differential equation:
\begin{equation}\label{phaseEquation}
\frac{\mathrm{d} \phi_i}{\mathrm{d}t}=k_\phi \: \exp \left \{ -\beta_\phi 
\left ( \frac{l_i}{2R_{\mathrm{c}}} - 1 \right ) \right \},
\end{equation}
where the positive constants $k_\phi$ and $\beta_\phi$ denote, respectively, the maximum rate of progress through the cell cycle and the extent to which progress is modified by mechanical compression and/or tension, while $l_i$ approximates the length of vessel element $i$: if vessel element $i$  has only two neighbours, elements $j$ and $k$, say, then  
\begin{equation}\label{celllength}
l_i=\frac{|\xx_i-\xx_k|+|\xx_i-\xx_j|}{2}.
\end{equation}
If element $i$ is a branching point, with more than two neighbours, then $l_i$ is the average over all distances between $i$ and its neighbours. 
Vessel element $i$ divides if $\phi_i=1$ and then the phases of the parent and new daughter cells are set to zero. 

The daughter cell is located randomly within the ball of radius $R_c$, centred on its parent and its fate is determined by its position and the degree of
mechanical compression being experienced by its parent (see Figure~\ref{fig:Sprouting} in main text). In more detail, we denote the position of the daughter cell relative to
its parent in polar coordinates by $\xx_{\mathrm{random}} = (r_{\mathrm{random}}, \alpha_{\mathrm{random}}, \gamma_{\mathrm{random}})$ where
$r_{\mathrm{random}} \in (0, R_c)$, $\alpha_{\mathrm{random}}\in [0,\pi]$ is the smallest angle between the vessel and the random vector (Figure~\ref{fig:VesselModel1}), and $\gamma_{\mathrm{random}}\in[0,\pi]$ is the angle between the plane spanned by the vessel elements connecting cells $i, j$ and $k$ and the random vector, 
$\xx_{\mathrm{random}}$. Then the new cell is a tip cell which forms a new capillary sprout if
\begin{equation}
(1-P_{\mathrm{sprout}})\left(\frac{\alpha}{2}+\gamma_{\mathrm{random}}\left(1-\frac{\alpha}{\pi}\right)\right)\leq\alpha_{\mathrm{random}}\leq (1+P_{\mathrm{sprout}})\left(\frac{\alpha}{2}+\gamma_{\mathrm{random}}\left(1-\frac{\alpha}{\pi}\right)\right)
\label{Prob_sprout}
\end{equation}
where
\begin{equation}
P_{\mathrm{sprout},i}=\frac{k_{\mathrm{spr}}}{K_{\mathrm{spr}}+(2R_{c}-l_i)_+},
\label{P_sprout}
\end{equation}
$l_i$ is defined by Equation (\ref{celllength}) and $(x)_+ =$ max ($0,x$).
The angles $\alpha_{\mathrm{random}}$ and $\alpha$ that appear in Equation (\ref{Prob_sprout}), and the probabilities with which sprouting or incorporation into the parent vessel happen, are illustrated in Figure~\ref{fig:Sprouting}.
Thus, $P_{\mathrm{sprout}}$ incorporates the influence that mechanical stretch of the has on the fate of its offspring, with less stretched vessel elements having smaller values
of $l_i$, larger values of $P_{\mathrm{sprout}}$ and, therefore, being more likely to form a new tip cell. 
In particular, if $\alpha_{\mathrm{random}}$ does not satisfy Equation (\ref{Prob_sprout}) then the new cell becomes a vessel element and contributes to elongation of the parent vessel.
In order to prevent rapid and non-smooth cell movement following proliferation, spring constants connecting each new cell/element to its neighbours are initially fixed at 
$0.1 S_c$ and then incremented by $0.1S_c$ over a short period (10 time-steps), until they reach $S_c$. 

For details regarding the computational implementation of our model and parameter values, we refer the reader to the Supplementary Materials, Sections S.1 and S.2.}

\begin{figure}[ht]
\begin{center}
\includegraphics[width=8.5cm]{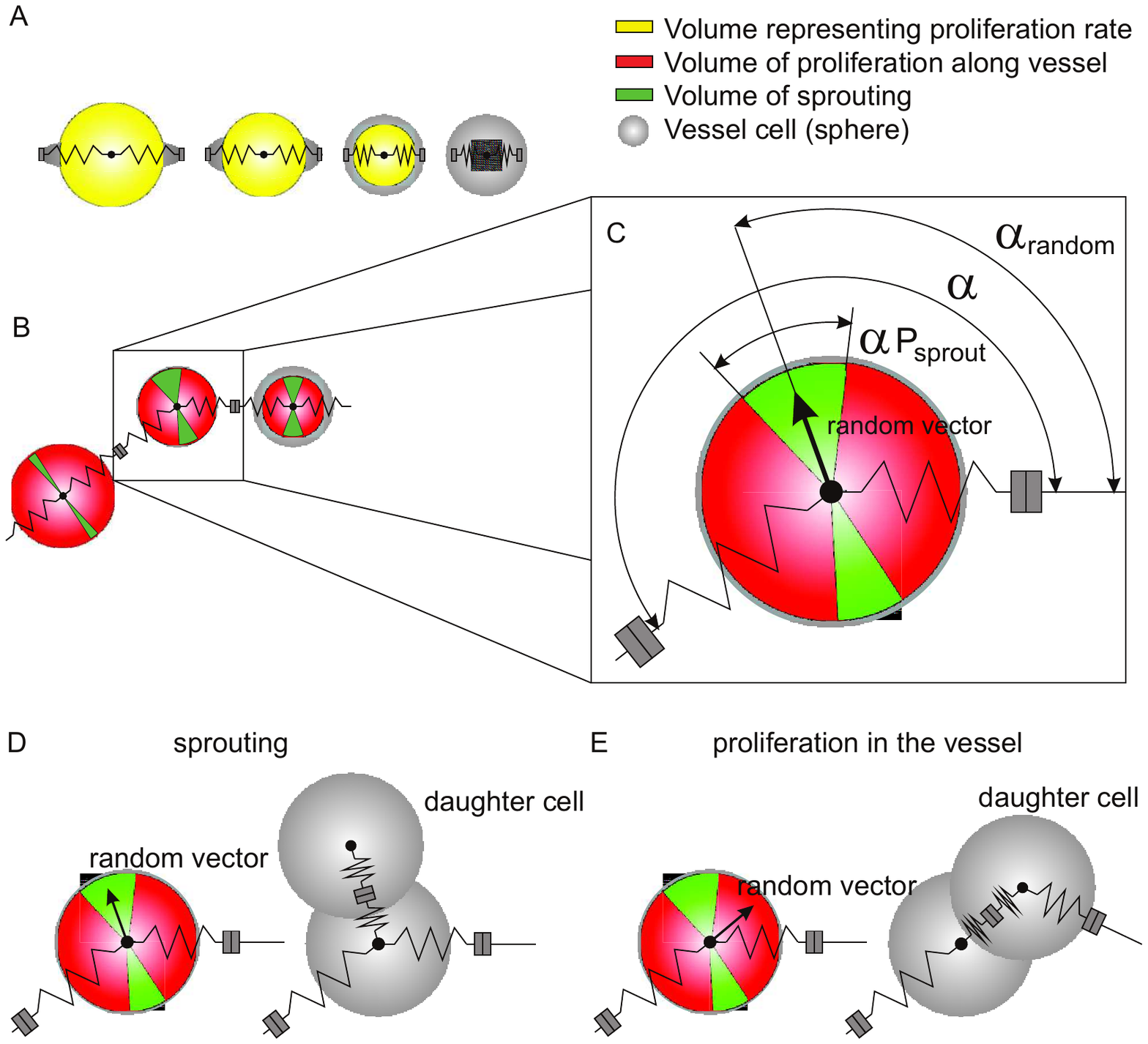}\\
\caption{{\bf Series of schematics showing how mechanical effects influence cell cycle progression and cell fate specification following cell division.} {\bf (A)} Mechanical stretch affects the rate of cell cycle progression of a vessel element. The diameters of the yellow circles are proportional to the rate at which the cell progresses through the cell cycle, larger circles indicating elongated cells which progress more rapidly through the cell-cycle. 
{\bf (B)--(C)}
When $\phi= 1$ a vessel cell divides, and the daughter cell is located at a randomly selected position 
$\xx_{\mathrm{random}} = (r_{\mathrm{random}}, \alpha_{\mathrm{random}}, \gamma_{\mathrm{random}})$ within the three-dimensional ball of radius $R_c$ centred on its parent cell. 
The fate of the daughter cell depends on the degree of stretch experienced by the parent cell, more compressed cells having greater probability of producing tip cells than mechanically stretched ones. The volume of the cells shaded red and green, respectively, indicate the probability that, when a cell divides, its offspring will be incorporated within the same vessel (red region) or form a new capillary sprout (green region), larger red regions corresponding to cells that are more stretched.
{\bf (D)--(E)} If the position of the daughter cell lies within the green volume, then it forms a new sprout  {\bf (D)}; 
otherwise, it is incorporated into the parent vessel {\bf (E)}. }
\label{fig:Sprouting}
\end{center}
\end{figure}

\subsection{Analysis and Visualisation of Simulation Results}
Since our model is stochastic, for each choice of parameter values, multiple simulations should be performed, using different random seeds, and suitable statistics extracted from the simulations to determine how the network structure evolves over time and how it is affected by changes in the parameter values. 

In order to characterise the vessel networks, and to facilitate future comparisons with experimental data, 
the following metrics are calculated for each simulation: (i) histograms showing the distribution of vessel element lengths, (ii) the total network length, (iii) the number of branches per unit length, (iv) the area covered by the network, (v) the displacement of the initial centre of mass and (vi) the tortuosity of the network (vessel tortuosity is determined by
decomposing the network into its component vessels and calculating 
the ratio of the sum of the lengths of all vessels to the sum of the distances between the endpoints of each vessel element, so that tortuosity $\geq 1$). 
Metrics (i)--(v) are calculated at fixed time points and the results aggregated to produce summary statistics showing how their mean and variance evolve over time (or as particular model parameters vary). We remark that a variety of metrics could be used to characterise the synthetic networks
that our model generates. For example, in a series of papers Konerding and coworkers \cite{konerding1999, konerding20013d, folarin2010} have measured inter-vessel distance, inter-branch distance, mean vessel diameter, vessel diameter and branching angle in three-dimensional corrosion casts of tumour networks. The five metrics we use are chosen for simplicity: we postpone
consideration of alternatives for future work.

\begin{figure}[ht!]
\begin{center}
\includegraphics[width=10.5cm]{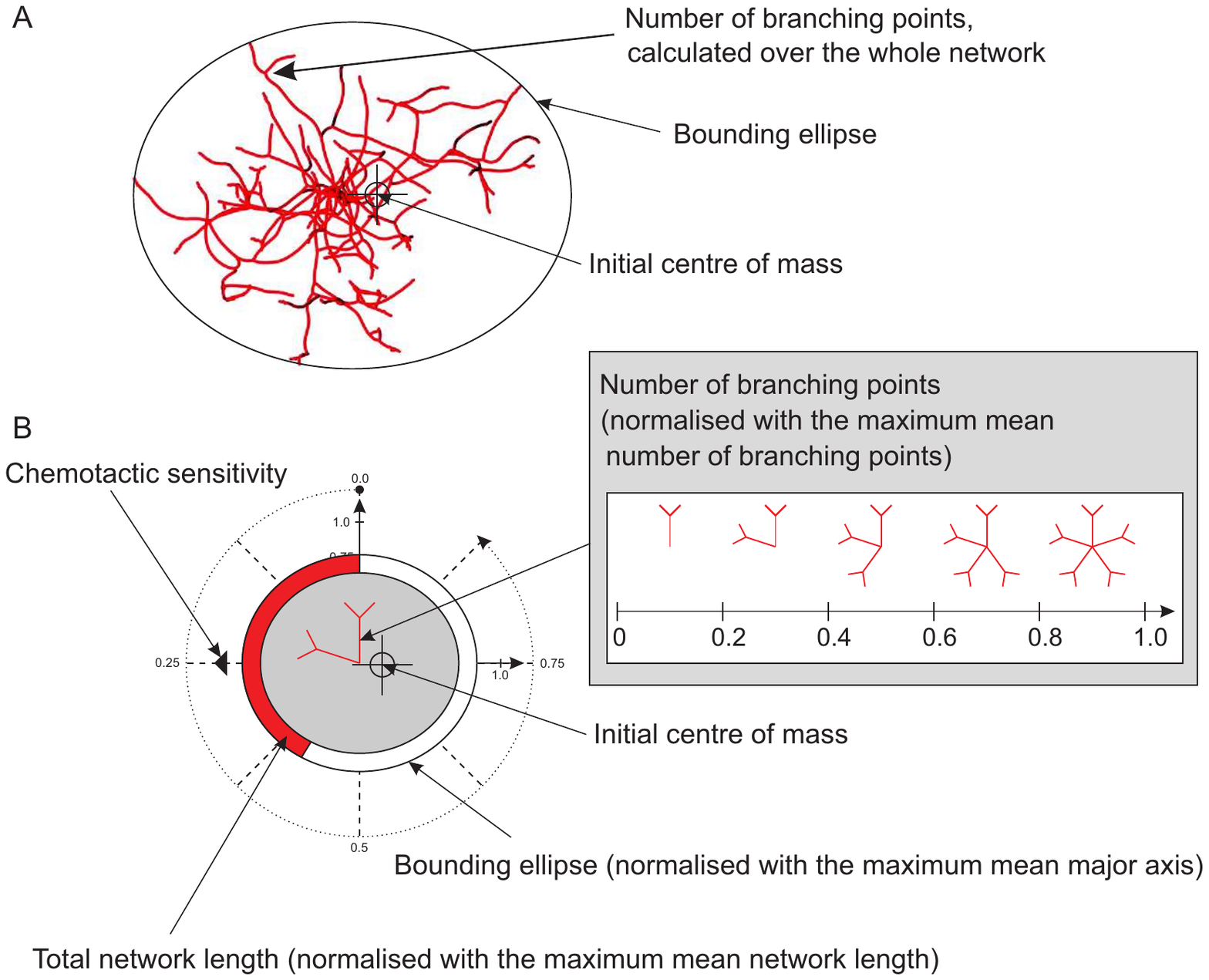}\\
\caption{{\bf Definition of glyphs used to visualise average network properties.} Glyphs summarise the mean properties of networks generated from multiple simulations of our model. (A) A typical simulated network. (B) A typical glyph. 
The area of the bounding ellipse is proportional to the smallest ellipse that circumscribes the network while the red pie chart that surrounds it is proportional to the total network length. The initial position of the network's centre of mass is indicated by a black hairpin while the number of branches on the red star at the centre of the ellipse provides a relative measure of the number of branching points. 
The chemotactic sensitivity of the tip cells is indicated by the size of the black triangle on the left-hand side (when this triangle is absent, no chemotactic force acts on the tip cells).}
\label{fig:GlyphExpl}
\end{center}
\end{figure}

We use two complementary approaches to visualise the results of our statistical analyses. When focussing on a single feature (e.g. total network length), we include plots showing how the mean and variance of that metric change over time (see Figure~\ref{fig:GlyphAndNetworkEvolution}) or as system parameters vary (see Figure~\ref{fig:GlyphNetworkSims1} and \ref{fig:GlyphNetworkSims2}) \cite{maguire2012taxonomy}. Since variation of model parameters may affect multiple metrics, we have designed a glyph so that we may visualise simultaneously, in a simple and comprehensive manner, several network metrics (Figure~\ref{fig:GlyphExpl}). In order to facilitate comparison of glyphs, they contain information on mean values only (where appropriate, information about standard deviations is presented elsewhere).
 
Figure \ref{fig:GlyphExpl} defines the glyphs that we use to visualise network properties. Each glyph has a bounding ellipse which is proportional to the mean bounding ellipse that circumscribes the vascular networks. 
{\color{black}
The main axes of the ellipse are determined by the difference between maximum and minimum values in $x$ and $y$ direction. One could also argue to take a bounding rectangle, but since the networks develops in a round way, an ellipse is more appropriate. Due to the isotropy of the growth process we only consider the bounding ellipse as a rough measure for the spatial extension.
}
The red pie chart that surrounds the ellipse is proportional to the total length of the network. The red star at the centre of the ellipse indicates the degree of branching of the network while the black hairpin indicates the initial (or default) position of the centre of mass of the network and indicates the extent to which the centre of mass of the network has been displaced during its evolution. The size of the black triangle on the left hand side indicates the chemotactic sensitivity of the network. 
We remark that chemotactic sensitivity and the displacement of the initial centre of mass are related: stronger chemotaxictic sensitivity will lead to larger displacement of the network's centre of mass.
As stated above, the main motivation for introducing glyphs is to facilitate comparison of networks generated using different parameter values. To effect these comparisons, the metrics that appear in the glyphs are normalised against values from a suitable series of reference simulations.

\section{Results}\label{sec:results}
In practice, vessel networks develop in heterogeneous environments and in response to multiple biophysical stimuli that include biochemical signals from angiogenic factors, such as VEGF, and mechanical forces from cell-cell and cell-substrate interactions. Integration of these signals by each cell regulates its phenotype (here, whether it is a tip cell or a vessel element), its rate of progress through the cell cycle, its movement and, thereby, the evolution of the entire network.
\textcolor{black}{In this paper we show how mechanical and chemical factors may influence the evolving morphology of vascular networks. 
We focus here on understanding the behaviour of the mathematical model and how simulation results involving multiple metrics can be presented in a clear and concise manner. Although our work is motivated by results obtained in \emph{in vitro} experiments, a detailed comparison with experimental data is beyond the scope of this work.} We perform sensitivity analysis to determine the influence of individual parameters on the network morphology. The impact of chemotaxis is investigated by varying the assumed constant chemotactic gradient, $c_x$ (for simplicity, we assume that the chemoattractant varies only in one direction, and that its gradient in
this direction is constant; see Supplementary Material, Equation (\ref{chemotacticForce})). The way in which mechanical stimuli may regulate endothelial cell proliferation and, thereby, vessel morphology, is studied by varying the mechanical sensitivity parameter $\beta_{\phi}$ that relates the rate of cell cycle progress to the degree of stretch that a cell is experiencing, larger values of $\beta_{\phi}$ indicating greater mechano-sensitivity (see Supplementary Material, Equation (\ref{phaseEquation})).
In what follows, unless otherwise stated, all simulations are generated using the default parameter values detailed in Table~\ref{tab:pars} and the initial network comprises two, straight and parallel vessels, each with three elements (tip-vessel-tip). 

\begin{figure}[ht!]
\begin{center}
\includegraphics[width=11.5cm]{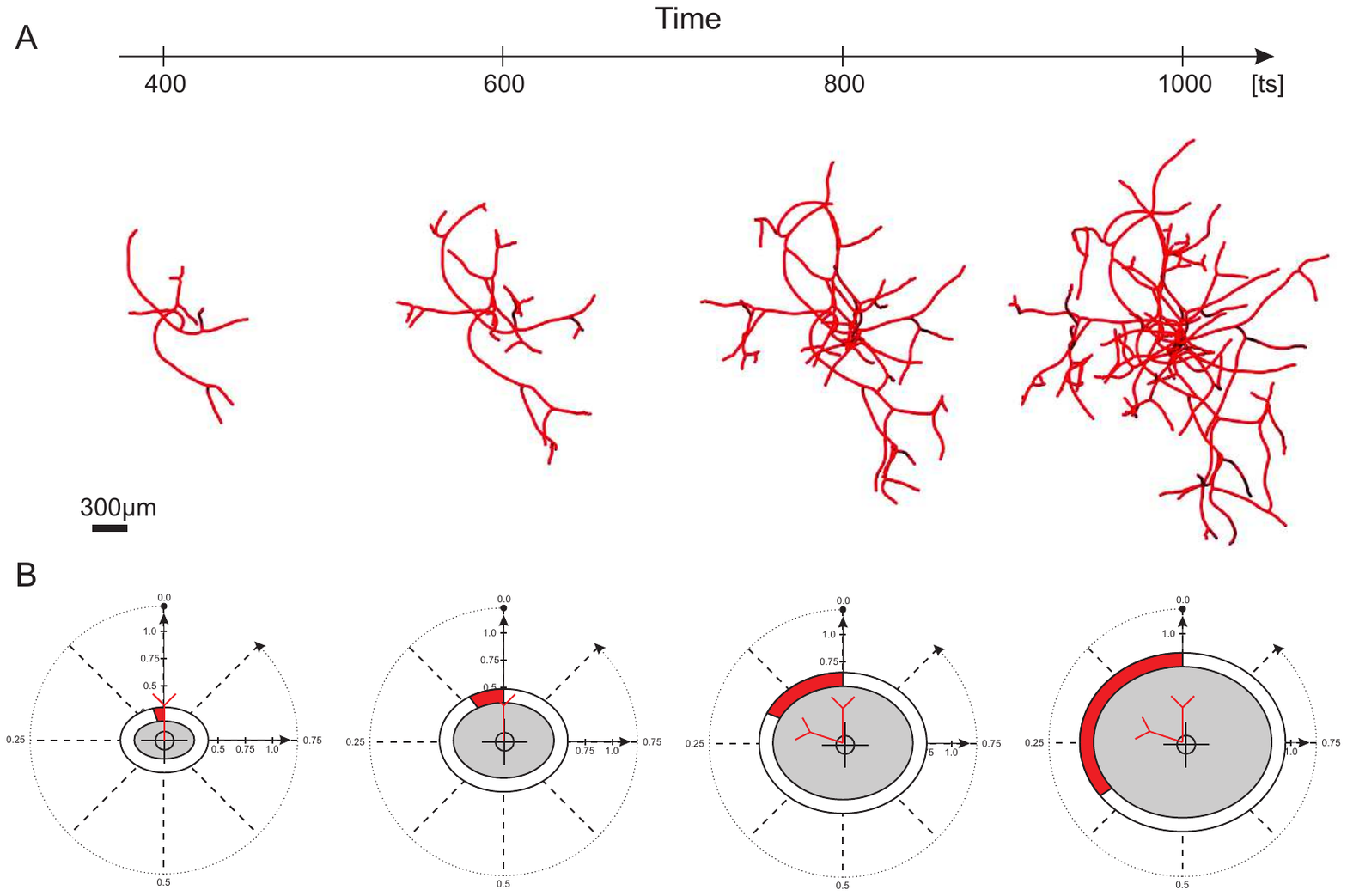}
\caption{{\bf Temporal network evolution, with associated glyphs.} 
{\bf (A)} Results from a typical dynamic simulation are presented at times $t=400, 600, 800, 1000$ timesteps (or $\mathrm{ts}$) and show how its size and structure change over time. 
{\bf (B)} Glyphs obtained by averaging 20 simulations indicate the average network properties at times $t=400, 600, 800, 1000\,\mathrm{ts}$.
Summary statistics showing how the mean and variance of other metrics evolve are presented in the Supplementary Material (see Figure \ref{fig:VesselLengthsTime8} ). 
Parameter values: as per Table S.1, except for $\chi$, the chemotactic sensitivity of the tip cells which is decreased from its default 
value of $\chi=0.1$ to $\chi = 0.0$.}
\label{fig:GlyphAndNetworkEvolution}
\end{center}
\end{figure}

Before presenting our parameter sensitivity analysis, we pause to consider a typical dynamic simulation in which chemotaxis is neglected.  
In Figure~\ref{fig:GlyphAndNetworkEvolution} we present results generated from a single realisation in which all parameters, except the chemotaxis coefficient, 
are fixed at their default values (see Supplementary Material, Table S.1). Over time, the network increases in size and develops an intricate pattern of interconnected vessel segments. The glyphs that accompany these images were obtained by averaging over 20 simulations. They reveal how the total vessel length and area spanned by the network increase over time and also how the average number of branches per unit length increases. For these simulations, since chemotaxis is inactive ($\chi=0$), the 
mean and variance of the network's centre of mass does not change significantly over time and nor do the mean and variance of the tortuosity. By contrast, the mean and variance in the total network length increase as the network expands, as does the mean number of branch points, although the variance
decreases over time (see Figure \ref{fig:VesselLengthsTime8} in Supplementary Material for information). 
We remark further that the simulation results presented in Figure \ref{fig:GlyphAndNetworkEvolution} suggest that realistic vascular networks can be generated in the absence of chemotaxis when mechanical stimuli alone drive cell proliferation and movement.

Having introduced typical network simulations and shown how glyphs can be used to characterise network features, we proceed to investigate how network morphology changes when model parameters vary. 
In the absence of detailed data with which to estimate system parameters, 
parameter ranges were chosen individually in order to obtain results that appear physically realistic (a comprehensive parameter sensitivity analysis is beyond the scope of the current work and, thus, we cannot guarantee that our choice of parameter values is the only feasible combination: we postpone
such considerations to future work). 
For example, large increases in the chemotactic sensitivity ($\chi$ in Equation (\ref{chemotacticForce})), or the sensitivity to random fluctuations ($\sigma$ in Equation (\ref{randomForce})),
can lead to "tip cell tear off." Here abnormally strong chemotactic forces cause tip cells to detach from the networks which then form degenerate clusters of vessel cells, because the directionality provided by the tip-cell is absent (results not shown). These findings highlight the important role that active tip-cell movement plays in network formation. 
We remark further that assigning a tip-cell fate to the end cell in a vessel segment following tear-off does not resolve this issue, leading instead to total disintegration of the network as new tip cells rapidly detach from the branch (results not shown). This (unphysical) behaviour is due to a mismatch between tip cell movement, vessel cell proliferation and the strength of cell-cell interactions, as represented by the spring constant (see Equation (\ref{mechanicalForce})).

\begin{figure}[ht!]
\begin{center}
\includegraphics[width=15.5cm]{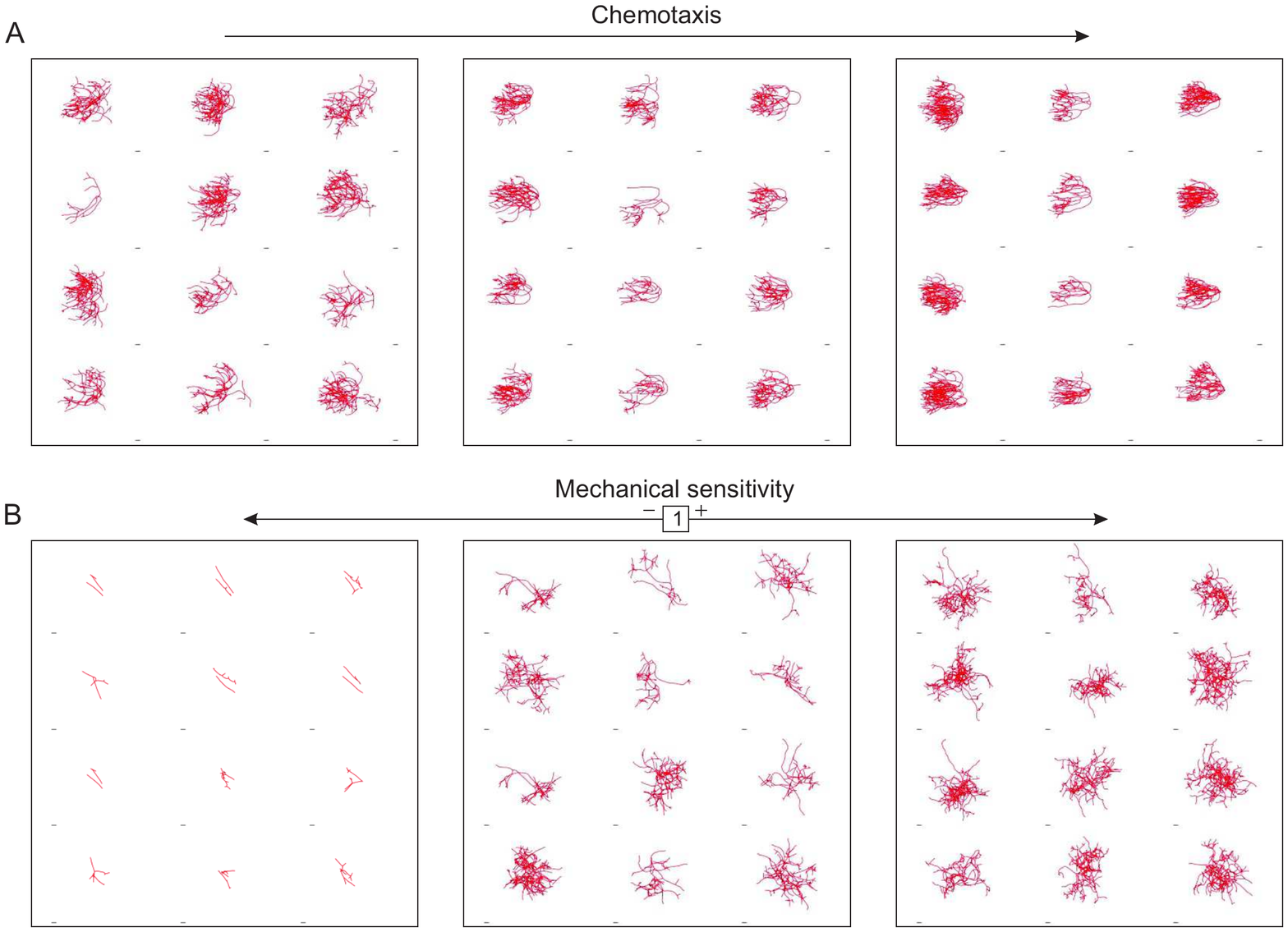}\\
\caption{{\bf Variability in simulation results for fixed parameter values and as individual parameter values vary.} 
For each choice of parameter values, we generate 12 simulations and output results at $t=900\,\mathrm{ts}$. 
{\bf (A)} As chemotactic sensitivity $\chi$ increases, the networks become more dense and oriented in the direction of the imposed chemotactic gradient
(right-to-left). 
{\bf (B)} As mechanical sensitivity $\beta_{\phi}$ increases, the network density and the average number of branches per unit vessel length increase.
Parameter values: as per Table~\ref{tab:pars}, except:
(A) $\beta_{\phi} = 100$, $\sigma = 0.4$, $k_{\mathrm{spr}} = 2 \times 10^{-4}$ and $\chi = 0.05, 0.10, 0.15$;
(B) $\chi = 0.0$, $\sigma = 0.4$, $k_{\mathrm{spr}} = 2 \times 10^{-4}$ and $\beta_{\phi} = 10, 100, 1000$.}
\label{fig:GlyphNetworkSims1}
\end{center}
\end{figure}

\begin{figure}[ht!]
\begin{center}
\includegraphics[width=15.5cm]{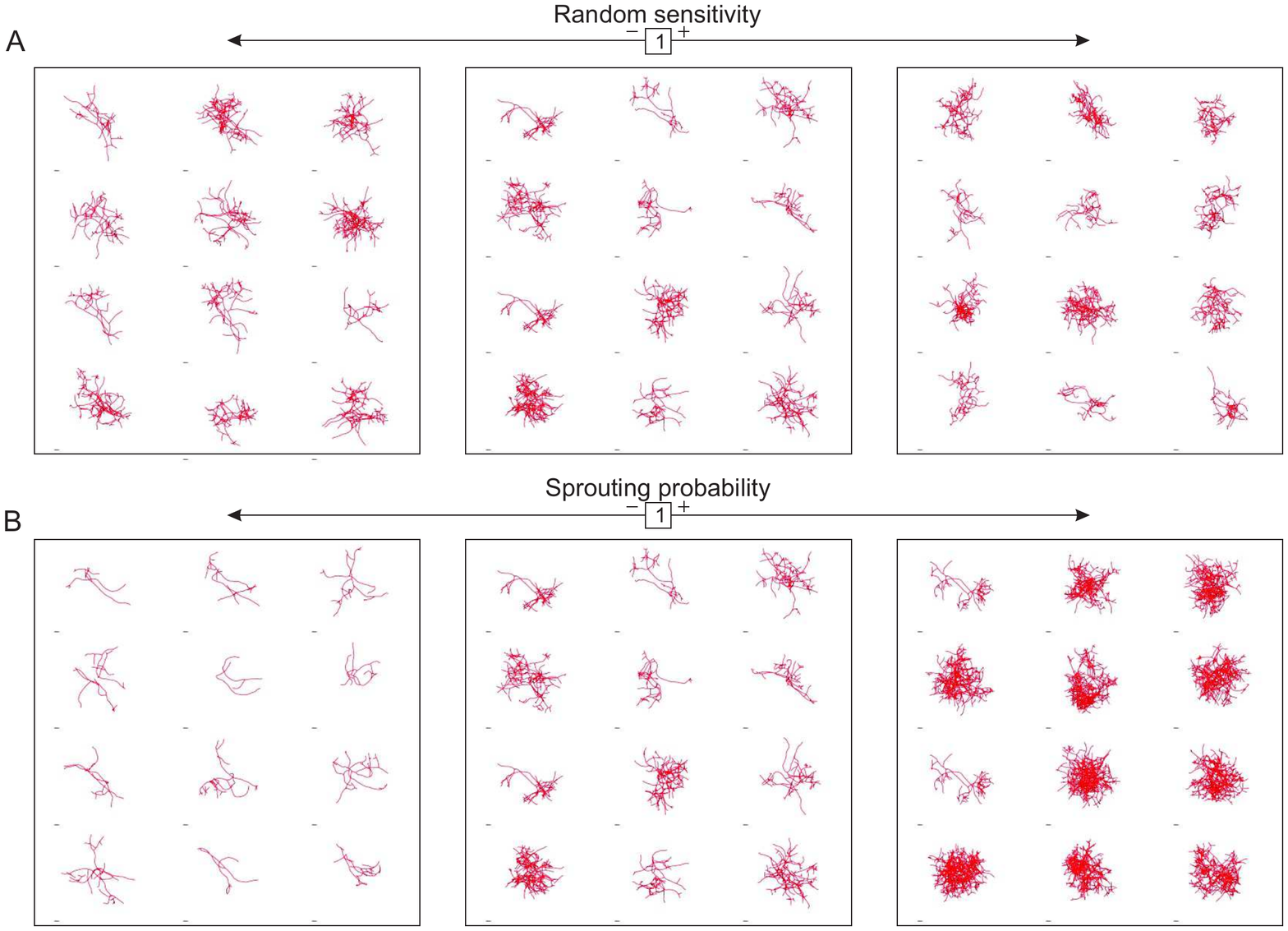}\\
\caption{{\bf Variability in simulation results for fixed parameter values and as individual parameter values vary.} 
For each choice of parameter values, we generate 12 simulations and output results at $t=900\,\mathrm{ts}$. 
{\bf (A)} Varying the cells' sensitivity to random fluctuations, $\sigma$, has a negligible effect on network size and morphology. 
{\bf (B)} Increasing the sprouting probability, $k_{\mathrm{spr}}$, increases the vascular density and average number of branch points per unit vessel length. 
Associated glyphs obtained by averaging over 70 simulations are presented in Figure~\ref{fig:Glyphnetwork}.
Parameter values: as per Table~\ref{tab:pars}, except:
(A) $\chi = 0.0$, $\beta_{\phi} = 100$, $k_{\mathrm{spr}} = 2 \times 10^{-4}$ and $\sigma = 0.2, 0.4, 0.8$;
(B) $\chi =0.0$, $\beta_{\phi} = 100$, $\sigma = 0.4$ and $k_{\mathrm{spr}} = 1 \times 10^{-4}, 2 \times 10^{-4}, 4 \times 10^{-4}$.}
\label{fig:GlyphNetworkSims2}
\end{center}
\end{figure}

\begin{figure}[ht!]
\begin{center}
\includegraphics[width=10.5cm]{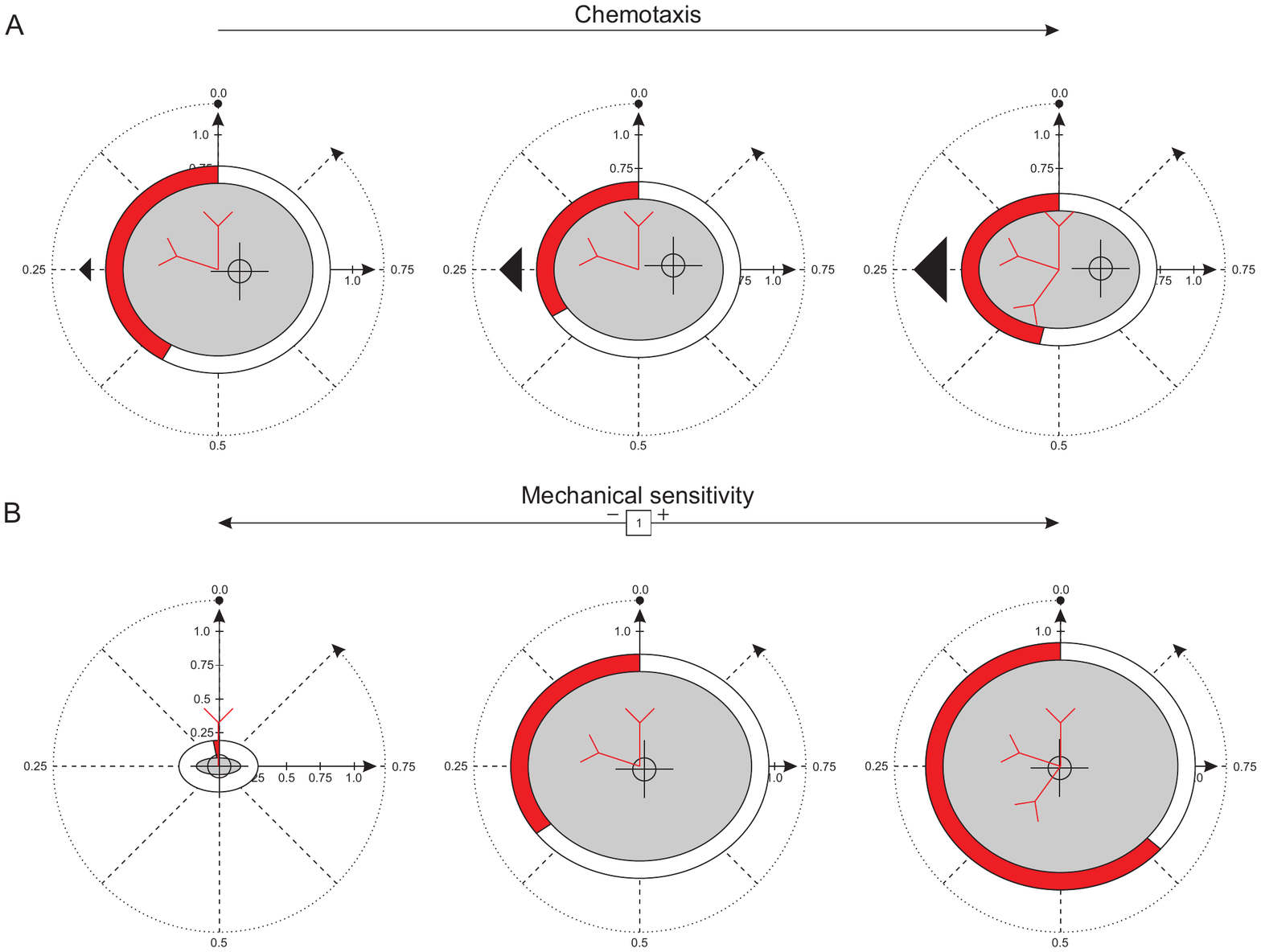}\\
\caption{{\bf Series of glyphs showing how chemotaxis and mechanical sensitivity affect network structure.} 
{\bf (A)} As the chemotactic sensitivity, $\chi$, of the tip cells increases, the network's centre of mass shifts to the right, it becomes more branched  but the total length of the network does not change significantly. 
{\bf (B)} As the mechanical sensitivity, $\beta_{\phi}$, of the vessel elements increases, the total length of the network and the average number of branches per unit vessel length increase. 
Each glyph depicts the mean behaviour of 70 vascular networks at $t = 900\,\mathrm{ts}$.
See Figure~\ref{fig:GlyphExpl} for an explanation of the glyphs. 
Parameter values: as per Table S.1, except:
(A) $\chi = 0.05, 0.10, 0.15$ and $\beta_{\phi} = 100$;
(B) $\beta_{\phi} = 10, 100, 1000$ and $\chi = 0.0$.}
\label{fig:GlyphnetworkCM}
\end{center}
\end{figure}

The results presented in Figures~\ref{fig:GlyphNetworkSims1},
\ref{fig:GlyphNetworkSims2} and \ref{fig:GlyphnetworkCM} 
illustrate the variability in size and morphology of the simulated networks for given sets of parameter values and as particular parameter values vary. 
In Figure~\ref{fig:GlyphNetworkSims1} we vary the chemotactic sensitivity, $\chi$, and the mechanical sensitivity, $\beta_{\phi}$, while in 
Figure~\ref{fig:GlyphNetworkSims2} we vary
the sensitivity to random fluctuations $\sigma$, and the sprouting probability $k_{\mathrm{spr}}$. For each set of parameter values, we present the structure of $12$ networks generated from different random seeds at $t=900$ timesteps ($\mathrm{ts}$) to illustrate the variability in the networks that can be generated from the same parameter values. 
 
As expected, the networks increase their directionality as the chemotactic sensitivity, $\chi$, increases, while varying the mechanical sensitivity, $\beta_\phi$, has a significant influence on the overall length of the networks and the number of branches per unit length of network. Equally, increasing $\sigma$, the parameter measuring the  cells' sensitivity to random fluctuations,  produces networks that are more tortuous while varying the branching probability, $k_{\mathrm{spr}}$, influences both the number of branches per unit length of the network and its total length.
(These results are reinforced and made more precise in the glyphs presented in Figures \ref{fig:GlyphnetworkCM} and \ref{fig:Glyphnetwork}.)
{\color{black}
When we vary both $\chi$ and $\beta_\phi$ we find that higher mechanical sensitivities lead to longer and more branched networks that cover a larger area while increasing the chemotactic strength decreases the area over which the network spreads, and increases the number of branches per unit length. The chemotactic strength strongly influences the centre of mass displacement (results not shown).}

\begin{table}[hptb]
\caption{{\bf Summary of parameter sensitivity analyses.} The results  presented in Figure~\ref{fig:VesselLengthsTime8} are
combined to show how changes in chemotactic sensitivity, $\chi$, mechanical sensitivity, $\beta_{\phi}$ and the sprouting probability, $k_{\mathrm{spr}}$, 
influence network metrics.  Parameter variations that elicit large changes in a particular metric are indicated by $+$; those that elicit no influence are indicated by $0$.}
\centering
\label{tab:influence} 
\begin{tabular}[t]{lcccc}
\hline\noalign{\smallskip}
 & length distr. & total length & branches per unit length & disp. of centre of mass \\[3pt]
chemo. sens.&0&0&0&+\\
mech. sens.&0&+&0&0\\
sprouting prob.&+&+&+&0\\
\noalign{\smallskip}\hline
\end{tabular}
\end{table}
The above parameter sensitivity analyses enable us to determine the influence of particular parameters on the structure of the emerging vessel network (see Table~\ref{tab:influence}). Not surprisingly, changes in the chemotactic sensitivity are predicted to shift the centre of mass of the vessel network towards the source of a chemoattractant. The mechanical sensitivity influences the temporal dynamics of network formation and the sprouting probability has a strong effect on the area spanned by the network area, the total network length and the number of branches per unit vessel length. Additionally, these results yield testable predictions about the impact on network morphology of varying particular system parameters.

Motivated by the qualitative comparisons that the glyphs provide and to facilitate future comparisons with experimental data, we now focus on more objective ways of quantifying the impact of parameter variations on network size and morphology. As mentioned in Section 2,  the following metrics are calculated for each simulation:
(i) histograms showing the distribution of vessel segment lengths, (ii) the total network length, (iii) the number of branches per unit length, (iv) the area spanned by the network, (v) the displacement of the initial centre of mass of the network, and (vi) the tortuosity of the network. 
For (B)--(E) the mean values are plotted as dots and the standard deviation as bars.

\begin{figure}[ht!]
\begin{center}
\includegraphics[width=11.5cm]{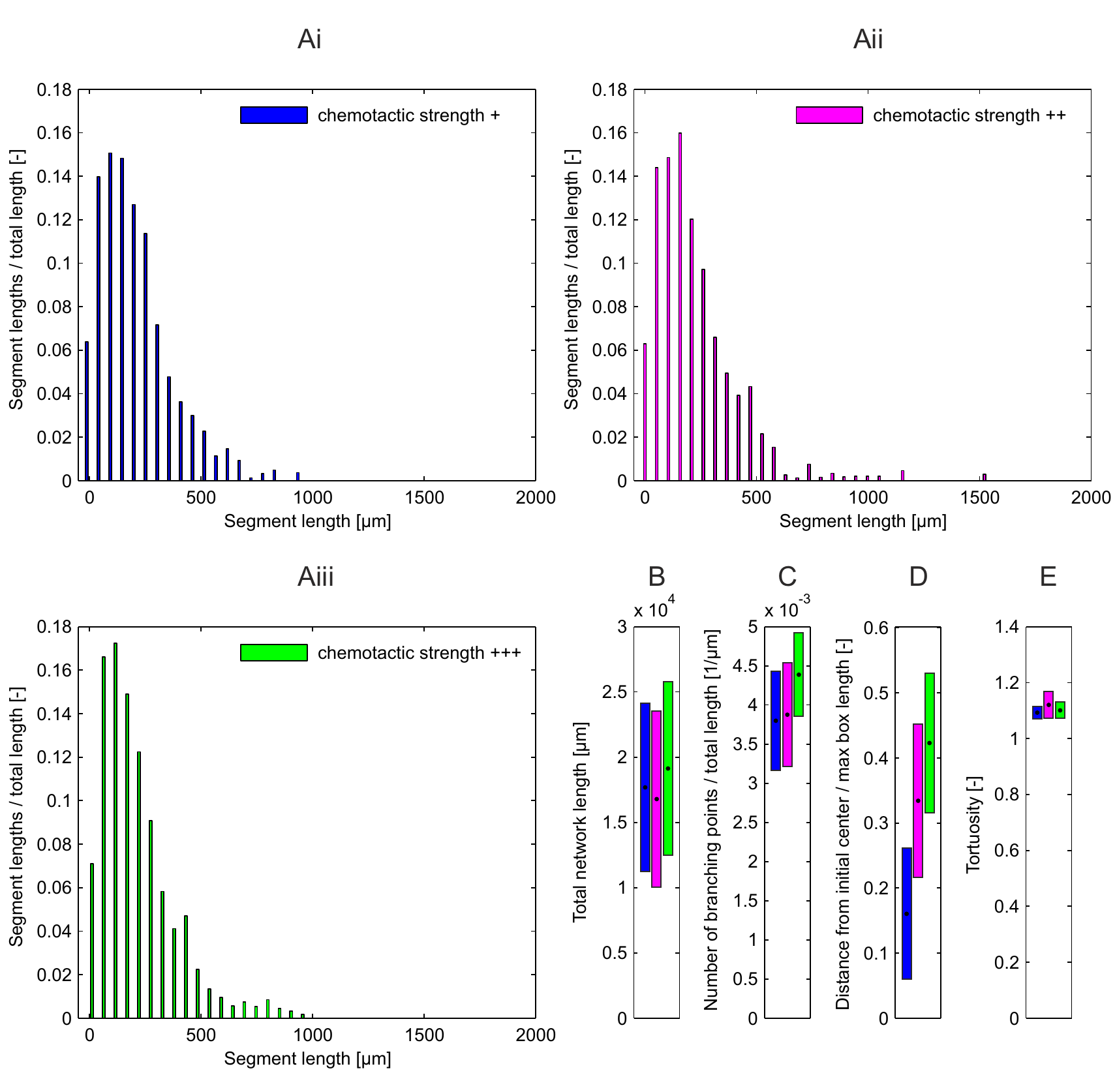}\\
\caption{{\bf Statistical analysis showing how network metrics depend on the chemotactic sensitivity, $\chi$.} 
As $\chi$ varies, only the position of the network's centre of mass at $t=900\,\mathrm{ts}$ shifts, in the direction of the chemoattractant gradient: 
changes in the other metrics are less pronounced.
{\bf (Ai--iii)} the distribution of vessel lengths for the different values of $\chi$; 
{\bf (B)} the total network length;
{\bf (C)} the number of branches per unit length;
{\bf (D)}, the displacement of the centre of mass; 
{\bf (E)} the tortuosity. Key: in (B)--(D), means illustrated by dots and standard deviations illustrated by bars were obtained by averaging over $70$ simulations.  
Parameter values: as per Table~\ref{tab:pars}, except $\chi = 0.05, 0.10, 0.15$. 
}
\label{fig:VesselLengths012}
\end{center}
\end{figure}

The influence of varying the chemotactic sensitivity, $\chi$, on the network morphology at $t=900\,\mathrm{ts}$ is depicted in Figure~\ref{fig:VesselLengths012} 
we choose this timepoint to ensure that any transients associated with the initial conditions
have disappeared and that metrics associated with the networks' internal structure (e.g. the average number of branch points per unit length or
tortuosity) have stabilised at constant values. 
While the distribution of segment lengths, the total network length and the number of branches per unit length do not change significantly, as expected, the centre of mass shifts significantly,  in the direction of the chemoattractant.
\begin{figure}[ht!]
\begin{center}
\includegraphics[width=11.5cm]{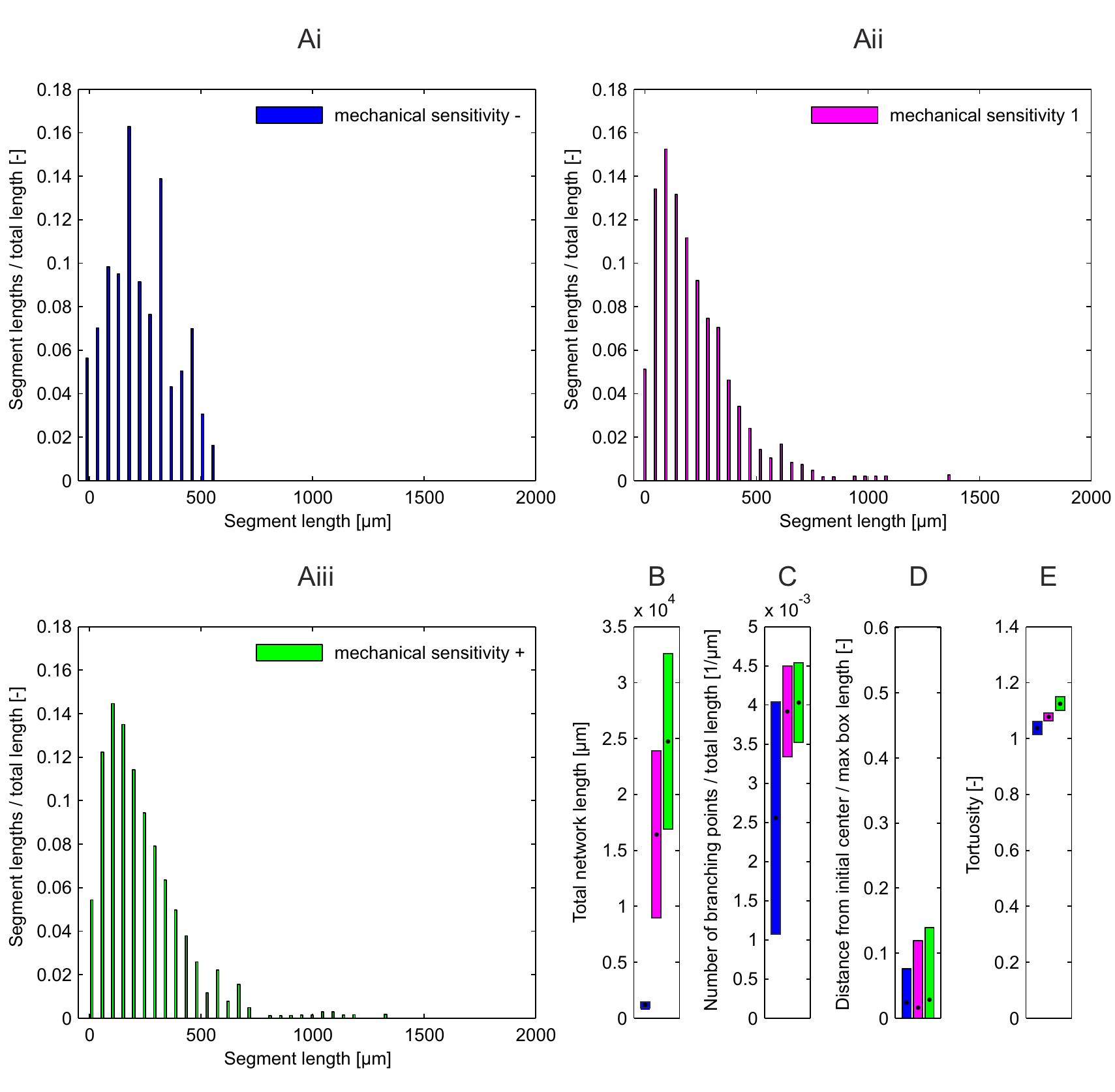}\\
\caption{{\bf Statistical analysis showing how network metrics depend on the mechanical sensitivity, $\beta_{\phi}$.} 
As $\beta_\phi$ varies, the total network length at $t=900\,\mathrm{ts}$ changes significantly; changes in the other metrics are less pronounced, although the tortuosity increases
slightly as $\beta_\phi$ increases.
{\bf (Ai--iii)} the distribution of vessel lengths for the different values of $\beta_\phi$; 
{\bf (B)} the total network length;
{\bf (C)} the number of branches per unit length;
{\bf (D)}, the displacement of the centre of mass; 
{\bf (E)} the tortuosity. Key: in (B)--(D), means illustrated by dots and standard deviations illustrated by bars were obtained by averaging over $70$ simulations.  
Parameter values: as per Table~\ref{tab:pars}, except $\beta_{\phi} = 10, 100, 1000$ and $\chi = 0.0$. 
}
\label{fig:VesselLengths384}
\end{center}
\end{figure}
Figure~\ref{fig:VesselLengths384} shows, in more detail than either Table 2 or Figure \ref{fig:GlyphnetworkCM}, that increasing the mechanical sensitivity, $\beta_\phi$, has a negligible effect on most network metrics, except for the total network length which is significantly larger.  
%
%
In addition to random fluctuations, drag and spring forces, vessel cells are also subject to an angular persistence, \textcolor{black}{which accounts for the characteristic \emph{persistence time} during which cells maintain memory of their former direction and therefore no significant changes of direction are observed \cite{selmeczi2005}. In this case, persistence is a consequence of the interaction with the surrounding tissue matrix} and suppresses small scale buckling instabilities (see Figures~ \ref{fig:AngularPersistence} and \ref{fig:NetworkStabilisation} in Supplementary Material). 
{\color{black}
Changing $\omega_a$, the strength of the angular persistence force, has a significant effect on vessel morphology, the stronger angular persistence producing networks which are less dense, have few branch points and are less tortuous (results not shown).} 
The statistics presented in Supplementary Figure \ref{fig:VesselLengthsTortuosity} show clearly that increasing $\omega_a$ skews the distribution of vessel lengths towards longer, straighter vessel segments, that it reduces tortuosity and the overall length of the network since the cells are more compressed in stiffer networks.  
Variation of the sprouting parameter $k_{\mathrm{spr}}$ (see Equations (\ref{Prob_sprout}) and (\ref{P_sprout})) has a similar effect on network morphology
(see Figure~\ref{fig:VesselLengths789}). Increasing $k_{\mathrm{spr}}$ shifts the distribution of segment lengths, leading, as we would expect, to more shorter vessel segments. The number of branches per unit length increases slightly and has a
marked effect on the overall length of the network.  In the absence of chemotaxis, the networks are isotropic and the tortuosity does not change as
$k_{\mathrm{spr}}$ varies.
\section{Discussion}\label{sec:discussion}

Understanding the mechanisms by which new blood vessels form has long fascinated and challenged experimental researchers, their interest stimulated in large part by the important roles played by angiogenesis and vasculogenesis during development, wound healing and their contributory roles in diseases such as rheumatoid arthritis, retinopathy of prematurity and cancer. Improvements in imaging techniques mean that it is now possible to collect high-resolution data showing how the spatial extent and morphology of vessel networks change over time and how the behaviour of the endothelial cells that form the networks is influenced by biophysical stimuli. 

In general terms, mathematical and computational modelling represent a natural framework for integrating diverse sources of experimental information 
about a biological system, for investigating the ways in which different processes may interact and generate observed behaviours, for testing hypotheses
and generating new predictions about the system's response to perturbations. In this paper we have developed a new, off-lattice model of vascular network formation. 
Where most existing hybrid models assume that chemotaxis is the dominant process driving endothelial cell proliferation and capillary sprout formation, our hybrid model allows us to investigate the possible roles played by mechanical stimuli. In particular, motivated by {\it in vitro} experiments performed by Liu et al. \cite{Liu:2007} and by Zheng et al. \cite{Zheng:2008}, we assume
that elongated cells increase their proliferation rate and that compressed cells are more likely to form new capillary sprouts when they
divide. Using these rules, our hybrid models generates vascular networks whose morphologies are qualitatively similar to those of real vascular networks cultivated {\it in vitro} (compare Figures~\ref{fig:MoffittImage} and \ref{fig:GlyphAndNetworkEvolution}).

We have performed extensive 3D numerical simulations to establish how different model parameters influence the size and morphology of the vascular networks. Since model simulations are stochastic, our parameter sensitivity analyses were based on network features
extracted from multiple simulations that were generated using different random seeds. The quantitative measures used to characterise the networks included the distribution of vessel lengths, the total network length, the average number of branches per unit length, tortuosity and the displacement of the network's centre of mass from its initial position. 
For example, increasing the chemotactic sensitivity of the tip cells does not appear significantly to affect the distribution of vessel lengths, the total network length or the average number of branches per unit length (see Figures~\ref{fig:GlyphnetworkCM}(A) and \ref{fig:VesselLengths012}), although it shifts the centre of mass in the direction of the chemoattractant. Equally, increasing the sensitivity of proliferating cells to mechanical effects leads to markedly larger networks (see Figures~\ref{fig:GlyphnetworkCM}(B) and
\ref{fig:VesselLengths384}). 
Increasing the sprouting probability has a significant effect on the area spanned by the network, total network length and the average number of branches per unit length (see Figures~\ref{fig:Glyphnetwork}(B) and \ref{fig:VesselLengths789}). Finally, 
increasing the random sensitivity or decreasing the strength of the angular persistence force 
(i.e., the stiffness of the gel in which the vessels are embedded) produces more tortuous networks (see 
Figures \ref{fig:Glyphnetwork}(A) and \ref{fig:VesselLengthsTortuosity}).  

There are many ways in which the work presented in this paper could be extended.
For example, by applying the statistical measures used to characterise the networks generated from our hybrid model to networks
generated using other approaches (e.g. cellular automata and Cellular Potts frameworks), it should be possible to identify those network features which are robust to the choice of cell-based framework and those which are not. 

\begin{figure}[ht!]
\begin{center}
\includegraphics[width=11.5cm]{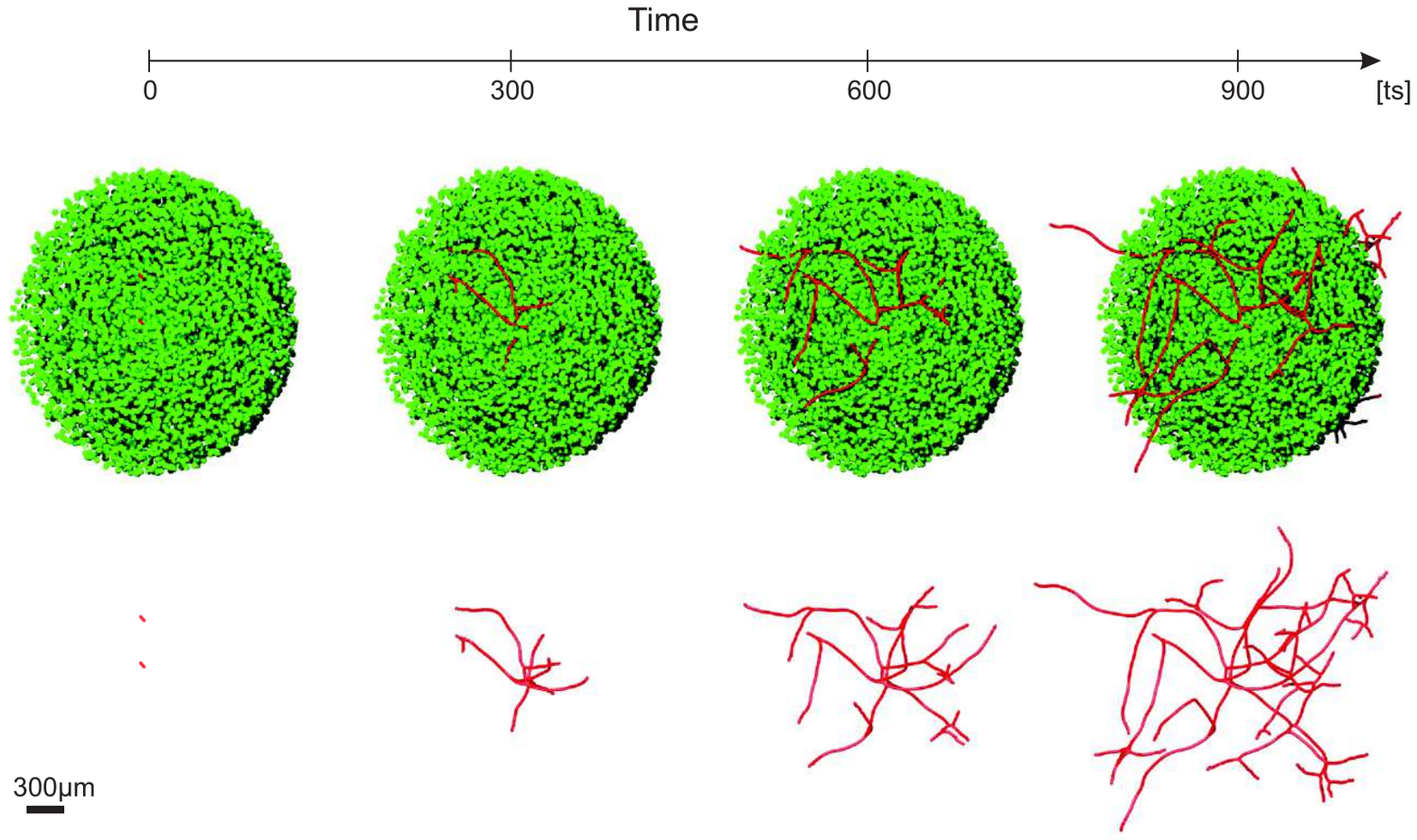}\\
\caption{{\bf Vasculogenesis in a 3D gel.} This figure illustrates how embedding the endothelial cells in a gel affects network evolution during the time period $0 \leq t \leq 900\,\mathrm{ts}$. The gel or host tissue (green particles) is modelled as a series of elastic spheres which interact mechanically with each other and the vessels (red particles). A cross-section through the spherical tissue is shown. Parameter values: as per Table~\ref{tab:pars}. For tumour cells only mechanical forces for compression are considered. }
\label{fig:VesselInTumour}
\end{center}
\end{figure}

A shortcoming of our hybrid model is the introduction of an angular persistence force that mimics the forces that endothelial cells experience due to interactions with their microenvironment and that dampen buckling instabilities induced by cell proliferation (see Figures~\ref{fig:AngularPersistence2},
\ref{fig:AngularPersistence} and \ref{fig:NetworkStabilisation} in Supplementary Material). \textcolor{black}{An alternative to the use of angular persistence to model the effect of the extracellular medium on network structure is explicitly to account for such interactions \cite{schlueter2012,lee2014,riching2014,schlueter2015}.} In Figure~\ref{fig:VesselInTumour} we present simulation results obtained by embedding the developing vessel network in a 3D gel which comprises non-proliferating, linearly elastic particles that possess the same mechanical properties as the endothelial cells. Force balances are applied to the vessel and gel elements, as before, except that forces due to cell-gel interactions supercede the angular persistence forces. The simulation results presented in Figure~\ref{fig:VesselInTumour} reveal that this model extension yields realistic network structures and could be used
to investigate how the mechanical properties of the gel or tissue in which the networks develop affect their size and morphology \cite{hahnfeldt1999tumor}. 

We have shown that mechano-transduction, \textcolor{black}{whose effects on solid tumour growth have been studied by Byrne and Preziosi \cite{Byrne2003} and Chaplain et al. \cite{chaplain2006},} is a key factor in our model, preventing compressed vessels from proliferating and stimulating sprout formation. Another important feature of our model is the active movement of tip cells that stabilises the network and establishes network morphology. Without tip cells, and/or a balance between tip-cell movement, vessel cell proliferation and cell-cell adhesion
(as modelled here via a spring constant), our simulated endothelial cells do not form a spatially distributed structure. Indeed, if the chemotactic sensitivity of the tip cells is too high, then they break away from the vessel cells which then form dense aggregates. 
Experimental studies have shown that tip- and vessel cells can change their phenotype \cite{guarani2011acetylation}, that such changes may be reversible, and may be regulated by subcellular signalling pathways \cite{blanco2012}.
In future work,  \textcolor{black}{by building upon previous work on multiscale modelling \cite{alarcon2005,ramisconde2008,owen2009,andasari2012,schlueter2015}}, we aim to increase the biological realism of our model by embedding, within the vessel cells, mathematical models of subcellular signalling pathways (e.g. delta-notch signalling \cite{bentley2009tipping}, vascular endothelial growth factor signalling
\cite{stefanini2012mmb,Zheng:2008} and Rac1 \cite{Liu:2007}) 
and using them to determine whether a particular cell adopts a tip- or vessel phenotype and/or whether a new daughter cell forms a new capillary sprout. 

In this paper, attention has focussed on characterising geometric features of the simulated networks: 
functional properties, such as average blood flow, wall shear stress and oxygen delivery rates, and their influence on
network evolution have been neglected.
Equally, the metabolic demands of the cells contained within the tissue, and their cross-talk with the evolving vessel networks, 
have been ignored.
Thus, in practice, our model focuses on the early stages of network development ({\it e.g} during vasculogenesis), before flow has been established, so that only stresses exerted on the vessels by the surrounding microenvironment need be considered. The impact of wall shear-stress and the metabolic demands of the perfused tissue can be incorporated in future model extensions that account for intravascular flow \cite{pries1998structural,alarcon2005b}, oxygen delivery and consumption 
\cite{alarcon2005,stephanou2005mathematical,owen2009,perfahl2011multiscale}.
We can then extract from these simulations similar statistics and construct glyphs which are similar
to those generated in this paper to investigate how inclusion of these processes affects the morphology of the resulting networks.
By further extending the model to account for the action of anti-angiogenic compounds such as endostatin or combretastin we could also
identify how such drugs should act in order to transiently normalise the tumour vasculature and, thereby, increase the tumour's sensitivity to radiotherapy
and standard chemotherapeutic agents \cite{jain2005,carmeliet2011}

 %
%

%
%


\begin{figure}[ht!]
\begin{center}
\includegraphics[width=11.5cm]{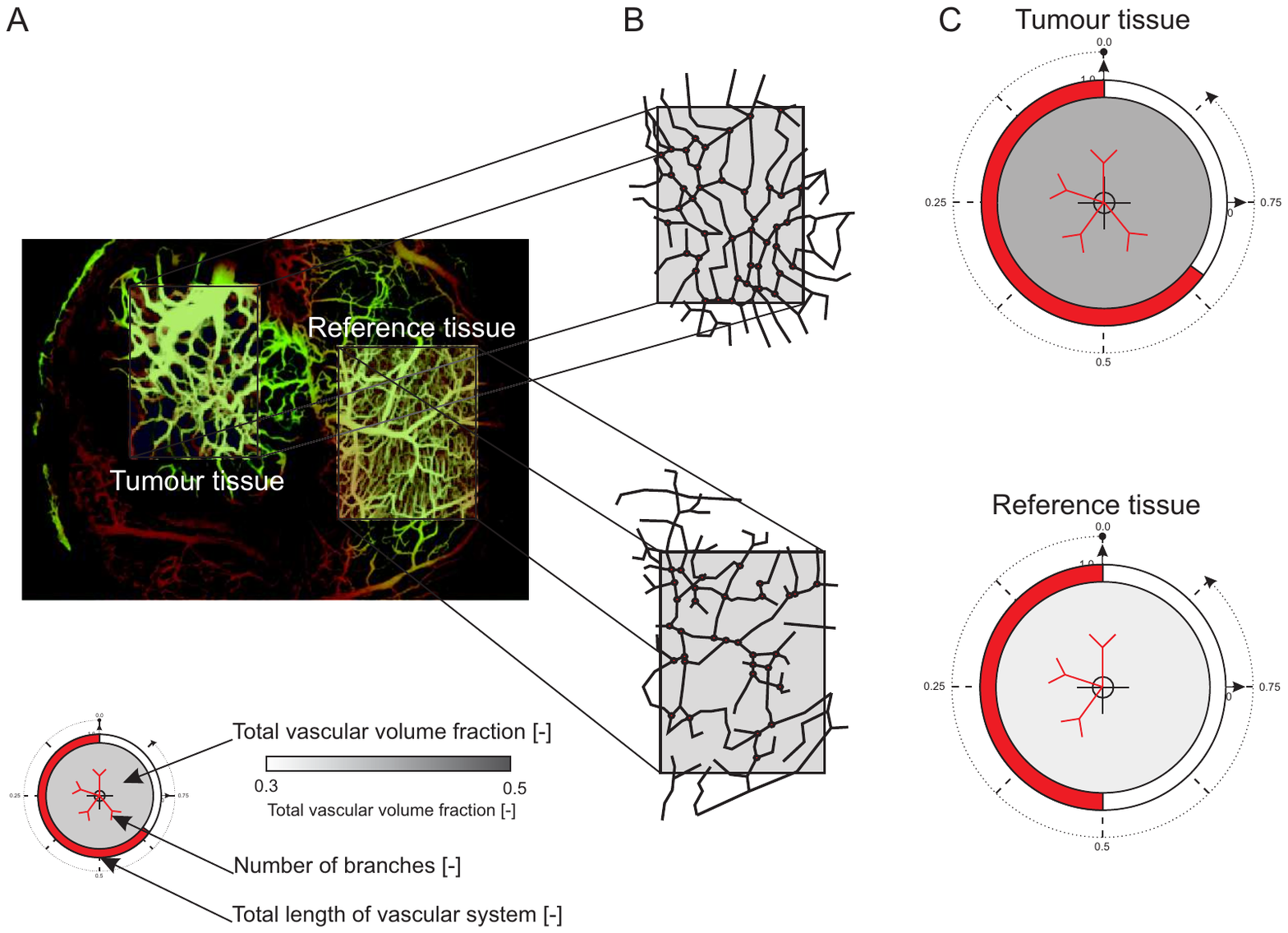}\\
\caption{{\bf Generating glyphs from experimental data.} These glyphs were generated using imaging data {\bf(A)} from cancerous and reference tissue areas \cite{steellaboratory}. The vascular graphs {\bf(B)} were extracted and used to construct glyphs. In contrast to the glyphs introduced before, the shading of the central circle indicates the vascular volume fraction in the region of interest. All glyph metrics were normalised so that the values for the reference (healthy) tissue are $0.5$.
}
\label{fig:GlyphFromExperiment}
\end{center}
\end{figure}
In our analysis of vascular networks, glyphs have served as objective tools with which to compare averaged properties of synthetic networks. They reveal how particular physical quantities change as the networks evolve over time, and show how model parameters and combinations thereof influence the network morphologies. The advantage of these glyphs lies in their high information content: they combine several network characteristics in a single image in an
objective and quantitative manner. In addition to characterising synthetically generated data sets, the glyphs could also be used to characterise {\it in vivo} images of vascular networks from animals and patients and, in so doing, provide a first step towards validation and parameterisation of angiogenesis and/or vasculogenesis.
As an illustrative example, in Figure~\ref{fig:GlyphFromExperiment} we present a high-resolution image from \cite{steellaboratory}  which shows two different tissue areas, one associated with reference, healthy tissue and the other associated with a tumour. Vascular graphs extracted from the two regions of interest are used to calculate the total length of the networks and the average number of branches per unit length. The corresponding glyphs are normalised so that the  properties associated with the reference tissue take half of their maximum values (and so that undervascularised necrotic areas and denser vascularised tissues can be easily identified). Our analysis reveals that, compared to the reference tissue, the tumour vasculature is longer, with thicker vessels that are more densely distributed across the tissue segment, and with more branch points per unit length of vessel. 
 
While there are many ways in which our hybrid model could be improved and extended, it demonstrates that mechanical regulation of endothelial cell proliferation and capillary sprout formation can generate realistic 3D vascular networks, even when chemotaxis is not active. Additionally, by introducing a range of statistical measures to characterise the networks and designing glyphs with which to visualise multiple network metrics, we are able to predict how network
properties evolve  over time and as key system parameters vary. Additionally, comparison of glyphs generated from synthetic data with those generated from 
{\it in vitro} and {\it in vivo} data reveal how, in the longer term and with suitable dynamic data, it may be possible to validate and parameterise our model and other hybrid models of vasculogenesis, angiogenesis and vascular remodelling.  

\paragraph*{Acknowledgements:}
This publication was based on work in part sponsored by the ``Federal Ministry of Education and Research'' Germany, funding initiative ``e:Med'' Systems Medicine (MultiscaleHCC) and funding initiative ``e:bio'' Systems Biology (PREDICT). 
TA gratefully acknowledges the Spanish Ministry of Economy (MINECO) for funding under grant MTM2015-71509-C2-1-R and Generalitat de Catalunya for funding under grant 2009SGR345. PKM was partially supported by the
National Cancer Institute, National Institutes of Health grant U54CA143970. MCL acknowledges funding from Moffitt PSOC NIH/NCI U54CA143970. BDH acknowedges funding from the Australian Research Council (DP110100795).
\bibliography{PerfahlBib}

\begin{thebibliography}{10}
\providecommand{\url}[1]{\texttt{#1}}
\providecommand{\urlprefix}{URL }
\expandafter\ifx\csname urlstyle\endcsname\relax
  \providecommand{\doi}[1]{doi:\discretionary{}{}{}#1}\else
  \providecommand{\doi}{doi:\discretionary{}{}{}\begingroup
  \urlstyle{rm}\Url}\fi
\providecommand{\bibAnnoteFile}[1]{%
  \IfFileExists{#1}{\begin{quotation}\noindent\textsc{Key:} #1\\
  \textsc{Annotation:}\ \input{#1}\end{quotation}}{}}
\providecommand{\bibAnnote}[2]{%
  \begin{quotation}\noindent\textsc{Key:} #1\\
  \textsc{Annotation:}\ #2\end{quotation}}
\providecommand{\eprint}[2][]{\url{#2}}

\bibitem{folkman1995angiogenesis}
Folkman J (1995) Angiogenesis in cancer, vascular, rheumatoid and other
  disease.
\newblock Nature Med 1: 27--30.
\bibAnnoteFile{folkman1995angiogenesis}

\bibitem{risau1997mechanisms}
Risau W (1997) Mechanisms of angiogenesis.
\newblock Nature 386: 671--674.
\bibAnnoteFile{risau1997mechanisms}

\bibitem{carmeliet2005angiogenesis}
Carmeliet P (2005) Angiogenesis in life, disease and medicine.
\newblock Nature 438: 932--936.
\bibAnnoteFile{carmeliet2005angiogenesis}

\bibitem{drake2003embryonic}
Drake C (2003) Embryonic and adult vasculogenesis.
\newblock Birth Defects Research Part C: Embryo Today: Reviews 69: 73--82.
\bibAnnoteFile{drake2003embryonic}

\bibitem{hahnfeldt1999tumor}
Hahnfeldt P, Panigrahy D, Folkman J, Hlatky L (1999) Tumor development under
  angiogenic signaling: a dynamical theory of tumor growth, treatment response,
  and postvascular dormancy.
\newblock Cancer Research 59: 4770--4775.
\bibAnnoteFile{hahnfeldt1999tumor}

\bibitem{Hubbard_2013}
Hubbard M, Byrne H (2013) Multiphase modelling of vascular tumour growth in two
  spatial dimensions.
\newblock J Theor Biol 29: 1015--1037.
\bibAnnoteFile{Hubbard_2013}

\bibitem{vilanova2013phasefield}
Vilanova G, Colominas I, Gomez H (2013) Capillary networks in tumour
  angiogenesis: from discrete endothelial cells to phase-field averaged
  descriptions via isogeometric analysis.
\newblock Int Jl Num Meth Biomed Eng 316: 70--89.
\bibAnnoteFile{vilanova2013phasefield}

\bibitem{arakelyan2003}
Arakelyan L, Merbl Y, Daugulis P, Ginosar Y, Vainstein V, et~al. Multi-scale
  analysis of angiogenic dynamics and therapy.
\newblock In: Preziosi L, editor, Cancer Modelling and Simulation, CRC Press,
  LLC, (UK).
\bibAnnoteFile{arakelyan2003}

\bibitem{balding1985mathematical}
Balding D, McElwain D (1985) A mathematical model of tumour-induced capillary
  growth.
\newblock J Theor Biol 114: 53--73.
\bibAnnoteFile{balding1985mathematical}

\bibitem{flegg2009three}
Flegg J, McElwain D, Byrne H, Turner I (2009) A three species model to simulate
  application of hyperbaric oxygen therapy to chronic wounds.
\newblock PLoS Comp Biol 5: e1000451.
\bibAnnoteFile{flegg2009three}

\bibitem{machado2011}
Machado M, Watson M, Devlin A, Chaplain M, McDougall S, et~al. (2011) Dynamics
  of angiogenesis during wound healing: a coupled in vivo and in silico study.
\newblock Microcirculation 18: 183--197.
\bibAnnoteFile{machado2011}

\bibitem{edgar2013}
Edgar L, Sibole S, Underwood C, Guilkey J, Weiss J (2013) A computational model
  of in vitro angiogenesis based on extracellular matrix fibre orientation.
\newblock Comp Meth Biomech Biomed Eng 16: 790--801.
\bibAnnoteFile{edgar2013}

\bibitem{manoussaki1996}
Manoussaki D, Lubkin S, Vernon R, Murray J (1996) A mechanical model for the
  formation of vascular networks in vitro.
\newblock Acta Biother 44: 271--282.
\bibAnnoteFile{manoussaki1996}

\bibitem{namy2004}
Namy P, Ohayon J, Tracqui P Critical conditions for pattern formation and in
  vitro tubulogenesis driven by cellular traction fields.
\newblock J theor Biol 227: 103--120.
\bibAnnoteFile{namy2004}

\bibitem{tosin2006}
Tosin A, Ambrosi D, Preziosi L Mechanics and chemotaxis in the morphogenesis of
  vascular networks.
\newblock Bull Math Biol 68: 1819--1836.
\bibAnnoteFile{tosin2006}

\bibitem{stephanou2015}
Stephanou A, Le~Floch S, Chauviere A A hybrid model to test the importance of
  mechanical cues driving cell migration in angiogenesis.
\newblock Math Model Nat Phenom 10: 142--166.
\bibAnnoteFile{stephanou2015}

\bibitem{Dyson2015}
Dyson R, Green J, Whiteley J, Byrne H (2015) An investigation of the influence
  of extracellular matrix anisotropy and cell-matrix interactions on tissue
  architecture.
\newblock J Math Biol 23: 1--35.
\bibAnnoteFile{Dyson2015}

\bibitem{stokes1991analysis}
Stokes C, Lauffenburger D (1991) Analysis of the roles of microvessel
  endothelial cell random motility and chemotaxis in angiogenesis.
\newblock J Theor Biol 152: 377--403.
\bibAnnoteFile{stokes1991analysis}

\bibitem{anderson1998}
Anderson A, Chaplain M (1998) Continuous and discrete mathematical models of
  tumour-induced angiogenesis.
\newblock Bull Math Biol 60: 857--899.
\bibAnnoteFile{anderson1998}

\bibitem{plank2004}
Plank MJ, Sleeman BD (2004) Lattice and non-lattice models of tumour
  angiogenesis.
\newblock Bull Math Biol 66: 1785--1819.
\bibAnnoteFile{plank2004}

\bibitem{mantzaris2004mathematical}
Mantzaris N, Webb S, Othmer H (2004) Mathematical modeling of tumor-induced
  angiogenesis.
\newblock J Math Biol 49: 111--187.
\bibAnnoteFile{mantzaris2004mathematical}

\bibitem{ambrosi2005review}
Ambrosi D, Bussolino F, Preziosi L (2005) A review of vasculogenesis models.
\newblock J Theor Med 6: 1--19.
\bibAnnoteFile{ambrosi2005review}

\bibitem{merks2009modeling}
Merks R, Koolwijk P (2009) Modeling morphogenesis in silico and in vitro:
  towards quantitative, predictive, cell-based modeling.
\newblock Math Model Nat Phenom 4: 149--171.
\bibAnnoteFile{merks2009modeling}

\bibitem{scianna2012review}
Scianna M, Bell C, Preziosi L (2013) A review of mathematical models for the
  formation of vascular networks.
\newblock J Theor Biol 333: 174--209.
\bibAnnoteFile{scianna2012review}

\bibitem{bentley2009tipping}
Blanco R, Gerhardt H (2013) Vegf and notch in tip and stalk cell selection.
\newblock Cold Spring Harb Perspect Med 3: a006569.
\bibAnnoteFile{bentley2009tipping}

\bibitem{merks2008contact}
Merks R, Perryn E, Shirinifard A, Glazier J (2008) Contact-inhibited chemotaxis
  in de novo and sprouting blood-vessel growth.
\newblock PLoS Comp Biol 4: e1000163.
\bibAnnoteFile{merks2008contact}

\bibitem{scianna2011multiscale}
Scianna M, Munaron L, Preziosi L (2011) A multiscale hybrid approach for
  vasculogenesis and related potential blocking therapies.
\newblock Prog Biophys Mol Biol 106(2): 450--62.
\bibAnnoteFile{scianna2011multiscale}

\bibitem{szabo2007network}
Szabo A, Perryn E, Czirok A (2007) Network formation of tissue cells via
  preferential attraction to elongated structures.
\newblock Phys Rev Lett 98: 38102.
\bibAnnoteFile{szabo2007network}

\bibitem{szabo2008multicellular}
Szabo A, Mehes E, Kosa E, Czirok A (2008) Multicellular sprouting in vitro.
\newblock Biophys J 95: 2702--2710.
\bibAnnoteFile{szabo2008multicellular}

\bibitem{szabo2012invasion}
Szab{\'o} A, Varga K, Garay T, Heged{\H{u}}s B, Czir{\'o}k A (2012) Invasion
  from a cell aggregate—the roles of active cell motion and mechanical
  equilibrium.
\newblock Phys Biol 9: 016010.
\bibAnnoteFile{szabo2012invasion}

\bibitem{van2013mechanical}
van Oers R, Rens E, LaValley D, Reinhart-King C, Merks R (2014) Mechanical
  cell-matrix feedback explains pairwise and collective endothelial cell
  behavior in vitro.
\newblock PLOS Comp Biol 10: e1003774.
\bibAnnoteFile{van2013mechanical}

\bibitem{stephanou2005mathematical}
Stephanou A, McDougall S, Anderson A, Chaplain M (2005) Mathematical modelling
  of flow in 2d and 3d vascular networks: applications to anti-angiogenic and
  chemotherapeutic drug strategies.
\newblock Math Comp Model 41: 1137--1156.
\bibAnnoteFile{stephanou2005mathematical}

\bibitem{watson2012dynamics}
Watson M, McDougall S, Chaplain M, Devlin A, Mitchell C (2012) Dynamics of
  angiogenesis during murine retinal development: a coupled in vivo and in
  silico study.
\newblock J Roy Soc Interface 13.
\bibAnnoteFile{watson2012dynamics}

\bibitem{alarcon2005}
Alarcon T, Byrne H, Maini P (2005) A multiple scale model for tumour growth.
\newblock Multiscale Model Simul 3: 440--475.
\bibAnnoteFile{alarcon2005}

\bibitem{owen2009}
Owen M, Alarc{\'o}n T, Maini P, Byrne H (2009) Angiogenesis and vascular
  remodelling in normal and cancerous tissues.
\newblock J Math Biol 58: 689--722.
\bibAnnoteFile{owen2009}

\bibitem{perfahl2011multiscale}
Perfahl H, Byrne H, Chen T, Estrella V, Alarc{\'o}n T, et~al. (2011) Multiscale
  modelling of vascular tumour growth in 3d: the roles of domain size and
  boundary conditions.
\newblock PLoS One 6: e14790.
\bibAnnoteFile{perfahl2011multiscale}

\bibitem{macklin2009}
Macklin P, McDougall S, Anderson ARA, Chaplain MAJ, Cristini V, et~al. (2009)
  Multiscale modelling and nonlinear simulation of vascular tumour growth.
\newblock J Math Biol 58: 765--798.
\bibAnnoteFile{macklin2009}

\bibitem{shirinifard2009}
Shirinifard A, Gens JS, Zaiden BL, Poplawski NJ, Swat M, et~al. (2009) 3d
  multi-cell simulation of tumour growth and angiogenesis.
\newblock PLoS One 4: e7190.
\bibAnnoteFile{shirinifard2009}

\bibitem{welter2008}
Welter M, Bartha K, Rieger H (2008) Emergent vascular network inhomogeneities
  and resulting blood flow patterns in a growing tumor.
\newblock J Theor Biol 250: 257--280.
\bibAnnoteFile{welter2008}

\bibitem{welter2009}
Welter M, Bartha K, Rieger H (2009) Vascular remodelling of an arterio-venous
  blood vessel network during solid tumour growth.
\newblock J Theor Biol 259: 405-422.
\bibAnnoteFile{welter2009}

\bibitem{hoehme2010prediction}
Hoehme S, Brulport M, Bauer A, Bedawy E, Schormann W, et~al. (2010) Prediction
  and validation of cell alignment along microvessels as order principle to
  restore tissue architecture in liver regeneration.
\newblock Proc Natl Acad Sci U S A 107: 10371--10376.
\bibAnnoteFile{hoehme2010prediction}

\bibitem{mclennan2012multiscale}
McLennan R, Dyson L, Prather K, Morrison J, Baker R, et~al. (2012) Multiscale
  mechanisms of cell migration during development: theory and experiment.
\newblock Development 139: 2935--2944.
\bibAnnoteFile{mclennan2012multiscale}

\bibitem{schaller2005multicellular}
Schaller G, Meyer-Hermann M (2005) Multicellular tumor spheroid in an
  off-lattice {V}oronoi-{D}elaunay cell model.
\newblock Phys Rev E 71: 051910.
\bibAnnoteFile{schaller2005multicellular}

\bibitem{pitt2009chaste}
Pitt-Francis J, Pathmanathan P, Bernabeu M, Bordas R, Cooper J, et~al. (2009)
  Chaste: a test-driven approach to software development for biological
  modelling.
\newblock Comp Phys Commun 180: 2452--2471.
\bibAnnoteFile{pitt2009chaste}

\bibitem{jackson2010cell}
Jackson T, Zheng X (2010) A cell-based model of endothelial cell migration,
  proliferation and maturation during corneal angiogenesis.
\newblock Bull Math Biol 72: 830--868.
\bibAnnoteFile{jackson2010cell}

\bibitem{Drasdo2009}
Drasdo D, Jagiella N, I~Ramis-Conde I, Vignon-Clementel I, Weens W (2010)
  Modeling steps from a benign tumor to an invasive cancer: examples of
  intrinsically multi-scale problems.
\newblock In: Chauviere A, Preziosi L, Verdier C, editors, Cell Mechanics: From
  Single Scale-Based Models to Multiscale Modeling, Chapman \& Hall/CRC. pp.
  379--417.
\bibAnnoteFile{Drasdo2009}

\bibitem{Liu:2007}
Liu W, Nelson C, Tan J, Chen C (2007) 3d microvascular architecture of
  pre-cancerous lesions and invasive carcinomas of the colon.
\newblock Circ Res 101: e44-e52.
\bibAnnoteFile{Liu:2007}

\bibitem{Zheng:2008}
Zheng W, Christensen L, Tomanek R (2008) Differential effects of cyclic and
  static stretch on coronary microvascular endothelial cell receptors and
  vasculogenic/angiogenic responses.
\newblock Am J Physiol Heart Circ Physiol 205: H794-H800.
\bibAnnoteFile{Zheng:2008}

\bibitem{maguire2012taxonomy}
Maguire E, Rocca-Serra P, Sansone S, Davies J, Chen M (2012) Taxonomy-based
  glyph design--with a case study on visualizing workflows of biological
  experiments.
\newblock IEEE T Vis Comp Gr 18: 2603--2612.
\bibAnnoteFile{maguire2012taxonomy}

\bibitem{hellstrom2007dll}
Hellstr{\"o}m M, Phng LK, Hofmann J, Wallgard E, Coultas L, et~al. (2007) Dll4
  signalling through notch1 regulates formation of tip cells during
  angiogenesis.
\newblock Nature 445: 776--780.
\bibAnnoteFile{hellstrom2007dll}

\bibitem{drasdo2005single}
Drasdo D, H{\"o}hme S (2005) A single-cell-based model of tumor growth in
  vitro: monolayers and spheroids.
\newblock Phys Biol 2: 133.
\bibAnnoteFile{drasdo2005single}

\bibitem{drasdo2001individual}
Drasdo D, Loeffler M (2001) Individual-based models to growth and folding in
  one-layered tissues: intestinal crypts and early development.
\newblock Nonlinear Analysis -- Theory Methods and Applications 47: 245--256.
\bibAnnoteFile{drasdo2001individual}

\bibitem{meineke2001cell}
Meineke F, Potten C, Loeffler M (2001) Cell migration and organization in the
  intestinal crypt using a lattice-free model.
\newblock Cell Prolif 34: 253--266.
\bibAnnoteFile{meineke2001cell}

\bibitem{walker2004epitheliome}
Walker D, Southgate J, Hill G, Holcombe M, Hose D, et~al. (2004) The
  epitheliome: agent-based modelling of the social behaviour of cells.
\newblock Biosystems 76: 89--100.
\bibAnnoteFile{walker2004epitheliome}

\bibitem{owen2011mathematical}
Owen M, Stamper I, Muthana M, Richardson G, Dobson J, et~al. (2011)
  Mathematical modeling predicts synergistic antitumor effects of combining a
  macrophage-based, hypoxia-targeted gene therapy with chemotherapy.
\newblock Cancer Res 71: 2826.
\bibAnnoteFile{owen2011mathematical}

\bibitem{konerding1999}
Konerding M, Malkush W, Klapthor B, van Ackem C, Fait E, et~al. (1999) Evidence
  for characteristic vascular patterns in solid tumours: quantitative studies
  using corrosion casts.
\newblock Brit J Cancer 80: 724--732.
\bibAnnoteFile{konerding1999}

\bibitem{konerding20013d}
Konerding M, Fait E, Gaumann A (2001) 3d microvascular architecture of
  pre-cancerous lesions and invasive carcinomas of the colon.
\newblock Brit J Cancer 84: 1354.
\bibAnnoteFile{konerding20013d}

\bibitem{folarin2010}
Folarin A, Konerding M, Timonen J, Nagl S, edley (2010) Evidence for
  characteristic vascular patterns in solid tumours: quantitative studies using
  corrosion casts.
\newblock Microvasc Res 80: 89--98.
\bibAnnoteFile{folarin2010}

\bibitem{selmeczi2005}
Selmeczi D, Mosler S, Hagedorn PH, Larsen NB, Flyvbjerg H (2005) Cell motility
  as persistent random motion: Theories from experiments.
\newblock Biophys J 89: 912--931.
\bibAnnoteFile{selmeczi2005}

\bibitem{schlueter2012}
Schlueter D, Ramis-Conde-I, Chaplain M (2012) Computational modeling of
  single-cell migration: The leading role of extracellular matrix fibers.
\newblock Biophys J 103: 1141--1151.
\bibAnnoteFile{schlueter2012}

\bibitem{lee2014}
Lee B, Zhou X, Riching K, Eliceiri K, Keely P, et~al. (2014) A
  three-dimensional computational model of collagen network mechanics.
\newblock PLoS One 9: e111896.
\bibAnnoteFile{lee2014}

\bibitem{riching2014}
Riching K, Cox B, Salick M, Pehlke C, Riching A, et~al. (2014) 3d collagen
  alignment limits protrusions to enhance breast cancer cell persistence.
\newblock Biophys J 107: 2546--2558.
\bibAnnoteFile{riching2014}

\bibitem{schlueter2015}
Schlueter D, Ramis-Conde I, Chaplain M (2015) Multi-scale modelling of the
  dynamics of cell colonies: insights into cell adhesion forces and cancer
  invasion from in silico simulations.
\newblock J R Soc Interface 12: 20141080.
\bibAnnoteFile{schlueter2015}

\bibitem{Byrne2003}
Byrne H, Preziosi L (2003) Modelling solid tumour growth using the theory of
  mixtures.
\newblock Math Med Biol 20: 341--366.
\bibAnnoteFile{Byrne2003}

\bibitem{chaplain2006}
Chaplain M, Graziano L, Preziosi L (2006) Mathematical modelling of the loss of
  tissue compression responsiveness and its role in solid tumour development.
\newblock Math Med Biol 23: 197--229.
\bibAnnoteFile{chaplain2006}

\bibitem{guarani2011acetylation}
Guarani V, Deflorian G, Franco C, Kr{\"u}ger M, Phng L, et~al. (2011)
  Acetylation-dependent regulation of endothelial notch signalling by the sirt1
  deacetylase.
\newblock Nature 473: 234--238.
\bibAnnoteFile{guarani2011acetylation}

\bibitem{blanco2012}
Bentley K, Mariggi G, Gerhardt H, Bates P (2009) Tipping the balance:
  robustness of tip cell selection, migration and fusion in angiogenesis.
\newblock PLoS Comp Biol 5: e1000549.
\bibAnnoteFile{blanco2012}

\bibitem{ramisconde2008}
Ramis-Conde I, Drasdo D, Chaplain M, Anderson A (2008) Modelling the influence
  of the e-cadherin - beta-catenin pathway in cancer cell invasion and tissue
  architecture: A multi-scale approach.
\newblock Biophys J 95: 155--165.
\bibAnnoteFile{ramisconde2008}

\bibitem{andasari2012}
Andasari V, Roper R, Swat M, Chaplain M (2012) Integrating intracellular
  dynamics using compucell3d and bionetsolver: Applications to multiscale
  modelling of cancer cell growth and invasion.
\newblock PLoS One 7: e33726.
\bibAnnoteFile{andasari2012}

\bibitem{stefanini2012mmb}
Stefanini M, Qutub A, MacGabhann F, Popel A (2012) Computational models of
  vegf-associated angiogenic processes in cancer.
\newblock Math Med Biol 29: 85--94.
\bibAnnoteFile{stefanini2012mmb}

\bibitem{pries1998structural}
Pries A, Secomb T, Gaehtgens P (1998) Structural adaptation and stability of
  microvascular networks: theory and simulations.
\newblock Amer J Physiol -- Heart and Circulatory Physiol 275: H349--H360.
\bibAnnoteFile{pries1998structural}

\bibitem{alarcon2005b}
Alarcon T, Byrne H, Maini P (2005) A design principle for vascular beds: the
  effects of complex blood rheology.
\newblock Microvasc Res 69: 156--172.
\bibAnnoteFile{alarcon2005b}

\bibitem{jain2005}
Jain R (2005) Normalisation of tumour vasculature: an emerging concept in
  antiangiogenic therapy.
\newblock Science 307: 58--62.
\bibAnnoteFile{jain2005}

\bibitem{carmeliet2011}
Carmeliet P, Jain R (2011) Principles and mechanisms of vessel normalisation
  for cancer and other angiogenic diseases.
\newblock Nature Reviews Drug Discovery 10: 417--427.
\bibAnnoteFile{carmeliet2011}

\bibitem{steellaboratory}
{D}epartment of~{R}adiation {O}ncology MGH (2013) {E}.{L}. {S}teel {L}aporatory
  {R}esearch {R}eport 2013.
\newblock {H}arvard {M}edical {S}chool.
\bibAnnoteFile{steellaboratory}

\bibitem{green2012}
Green S (2012).
\newblock Particle simulation using {C}{U}{D}{A}.
\newblock Website, CUDA documentation.
\newblock
  \texttt{http://docs.nvidia.com/cuda/samples/5\_Simulations/particles/doc/particles.pdf}.
\bibAnnoteFile{green2012}

\end{thebibliography}

\newpage
\renewcommand{\thepage}{S\arabic{page}}
\setcounter{page}{1}
\renewcommand{\thesection}{S.\arabic{section}}  
\setcounter{section}{0}
\renewcommand{\thesubsection}{\thesection.\arabic{subsection}}  
\setcounter{subsection}{0}
\renewcommand{\thetable}{S.\arabic{table}}   
\setcounter{table}{0}
\renewcommand{\thefigure}{S.\arabic{figure}}
\setcounter{figure}{0}
\renewcommand{\theequation}{S.\arabic{equation}}
\setcounter{equation}{0}
%
\section*{Supplementary Material}\label{SuppMat}
The Supplementary Material comprises three sections in which we describe  \textcolor{black}{the computational algorithm and parameter values that were used to generate the numerical results (S.1 and S.2, respectively), before presenting additional simulation results (S.3). }

\section{The Computational Algorithm}
\begin{figure}[ht!]
\begin{center}
\includegraphics[width=10.5cm]{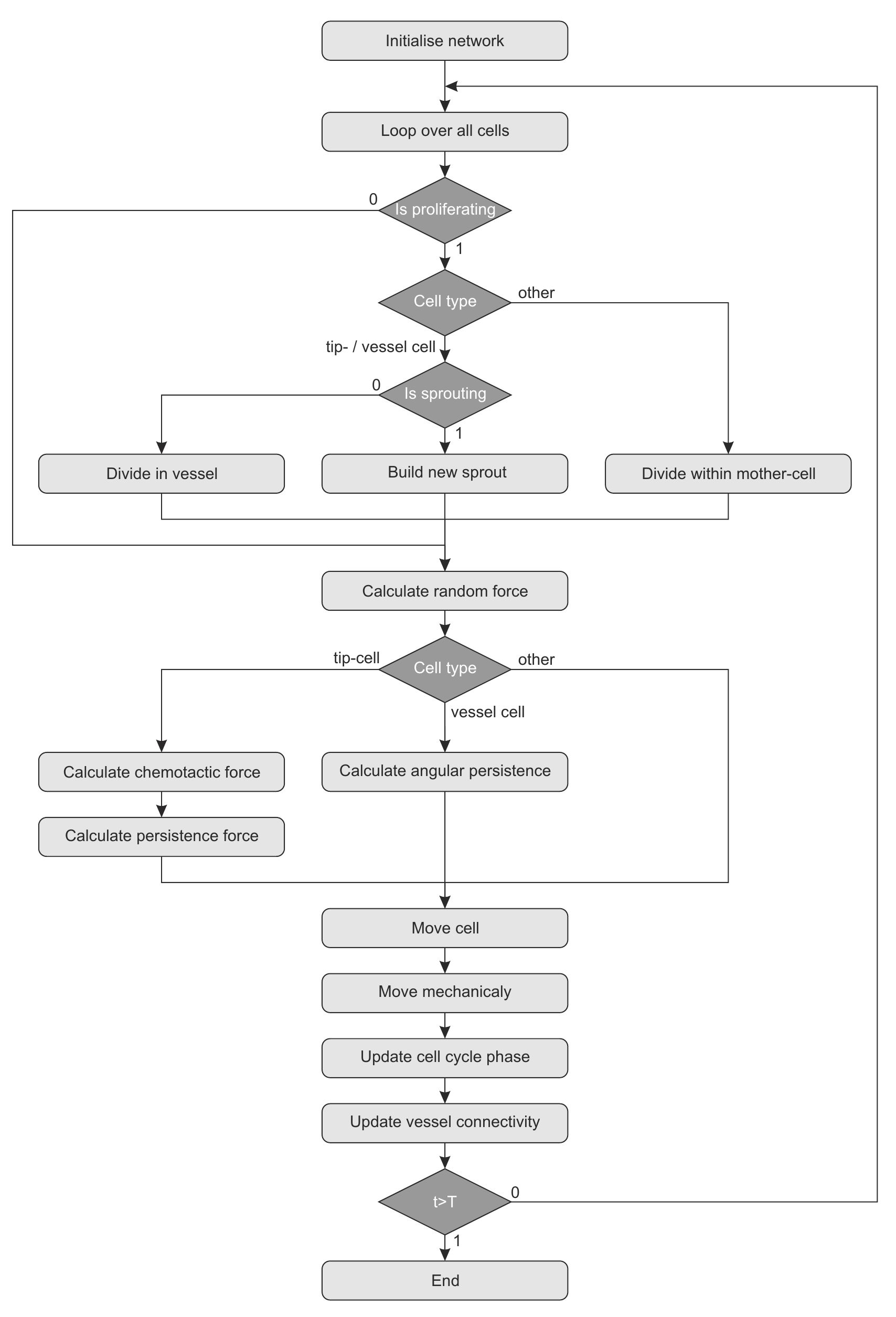}\\
\caption{{\bf Computational algorithm.} This figure shows the computational algorithm used to simulate the off-lattice model of vasculogenesis.
}
\label{fig:Algorithm}
\end{center}
\end{figure}

We now outline the algorithm that is used to generate numerical simulations of our hybrid, agent-based model (see, also, Figure \ref{fig:Algorithm}).
We consider a three-dimensional Cartesian domain, of size $W_X \times W_Y \times W_Z$, discretised uniformly, with spacings $\Delta X$, $\Delta Y$ and $\Delta Z$ in the $x$, $y$ and $z$ directions, respectively. While an off-lattice approach is used to model the evolution of the vascular network, the domain discretisation is used to sort the cells (to increase computational efficacy in finding interacting neighbours \cite{green2012}) and, thereby, to increase the speed of the numerical simulations. As stated above, the total number of segments (both tip and vessel segments) at time $t$ is denoted by $N=N(t)$. The vascular network structure is stored in an undirected graph in which cell centre coordinates $\mathcal{N}$ and cell-cell connections $\mathcal{E}$ are recorded. The adjacency matrix $\mathcal{E}$ contains 2-tuples of the node numbers of all connections in the vessel network.

\begin{enumerate}
\item {\bf Initialise the simulation ($t=0$)}\\
The initial network is defined by specifying a fixed number of vessels, the number of segments within each vessel, their positions and connectivity. Unless otherwise stated, the initial network comprises two parallel vessels, each formed of three elements (tip-vessel-tip). The initial cell cycle phase of each vessel element and the chemoattractant gradient are also prescribed ($\nabla c = (-c_x,0,0)$).
\item {\bf Increment time ($t \rightarrow t + \Delta t$) and loop over all cells ($i=1, \ldots, N(t)$)}\\
The following steps are performed for each element in the network. 
\begin{enumerate}
\item {\bf Test for cell proliferation (vessel elements only) }\\
If element $i$ is a vessel element and $\phi_i \geq 1$ then it proliferates and its daughter cell is placed at a random position within a sphere of radius $R_c$ centred on the parent cell. The daughter cell forms a new capillary sprout if its location is such that
inequalities (\ref{Prob_sprout}) are satisfied; otherwise, it becomes a vessel cell and contributes to elongation of the parent vessel.  
\item {\bf Update the position of element $i$}\\
Different forces act on vessel and tip cells (see Equations (\ref{tip:movement}) and (\ref{vessel:movement})).  
We assume that these forces act on two different time-scales: a slow time-scale is associated with chemotactic and random movement, and a faster one with mechanically induced cell movement. We exploit this separation of time-scales by using a two-step method to integrate the equations of motion and update $\xx_i$, the position of element $i$. 
\begin{itemize}
\item {\bf Step 1.}
Calculate the random force $\FF_i^r$ using Equation~(\ref{randomForce}). If cell $i$ is a tip-cell, then 
calculate the chemotactic  and persistence forces $\FF_i^c$ and $\FF_i^p$ using Equations~(\ref{chemotacticForce}) and (\ref{persistenceForce}).  
Update $\xx_i$, the position of cell $i$, using an explicit Eulerian method to integrate Equation (\ref{tip:movement}) for tip cells or (\ref{vessel:movement}) for vessel cells, and neglecting the mechanical and angular persistence forces. 
\item {\bf Step 2.}
Calculate the mechanical force $\FF_i^m$ using Equation~(\ref{mechanicalForce}) and if cell $i$ is a vessel cell
then calculate the angular persistence force $\FF_i^a$ using Equation~(\ref{angularpersistenceForce}). 
If element $i$ is a new cell, then its spring constant is increased in increments of $0.1 S_c$ on consecutive time steps until it reaches the value $S_c$. 
Then  $\xx_i$, the position of cell $i$, is updated using an explicit Eulerian method to integrate Equation (\ref{tip:movement}) for tip cells or (\ref{vessel:movement}) for vessel cells. This step is repeated until the displacement on consecutive steps is less than a threshold
$\varepsilon$ or  the number of iterations has reached the maximum value, $\max_{\mathrm{iter}}$.
\end{itemize}
\item {\bf Update the cell cycle phase of cell $i$ using Equation (\ref{phaseEquation}) (vessel elements only) }
\item {\bf Update vessel connectivity}\\
New connections are added to the vascular graph, and old ones removed when tip cells anastamose with other vessel segments or detach 
from their branch (detachment occurs if a tip-cell moves beyond the cut-off distance $l_c$, see Equation~(\ref{mechanicalForce})).
\end{enumerate}
\item If $t \geq T$ then end the simulation; otherwise, return to step 2.
\end{enumerate}
A flowchart of the computational algorithm is given in Figure~\ref{fig:Algorithm}. The algorithm illustrates how other species (e.g. normal cells, tumour cells, immune cells and tissue matrix) could be incorporated into the computational framework. We remark further that the order in which the algorithmic steps 
are carried out is arbitrary; reordering generates identical, mean results (results not included).

\section{The Default Parameter Values}
In the table that follows we state the default parameter values that were used to generate the numerical simulations. We stress that these parameter values were chosen to generate physically realistic results and not by fitting the model to experimental data (an investigation is beyond the scope of the current work). 

\begin{table}[hptb]
\caption{Summary of the parameters used in the computational model of vasculogenesis, along with their physical meaning and estimates of their dimensionless values (lengths are scaled with $\mu m$ and times with $sec$).}
\centering
\label{tab:pars} 
\begin{tabular}[t]{lll}
\hline\noalign{\smallskip}
Parameter & Physical meaning & Default value \\[3pt]
$R_{c}$ & target cell radius & $5$\\
$l_c$ & cell interaction radius & $3R_c$\\
$\mu $ & drag coefficient & $1$\\
$\sigma$ & sensitivity to random fluctuations &  $0.4$\\
$S_c$ &spring constant for compressed cells & $5.56\times 10^{-6}$\\
$100S_c$&spring constant for stretched cells & $5.56\times 10^{-4}$\\
$\chi$ & chemotactic sensitivity & $0.1$\\
$c_x$ & chemoattractant gradient& $5.56\times 10^{-4}$\\
$w_p$ & directional persistence coefficient & $0.4$\\
$\tau$ & directional persistence time & $180$\\
$\omega_a$ & angular persistence spring constant& $5.56\times 10^{-5}$\\
$k_{\phi}$ & progression rate through the cell cycle&  $1.39\times 10^{-5}$\\
$\beta_\phi$ & cell cycle progression sensitivity to cell elongation and compression & $100.0$\\
$k_{\mathrm{spr}}$ & probability of sprouting & $2.0\times10^{-4}$\\
$K_{\mathrm{spr}}$ & parameter for sprouting probability& $0.01$\\
\hline
$W_X$ & domain length ($x-$direction) & $1000$\\
$W_Y$ & domain length ($y-$direction) & $1000$\\
$W_Z$ & domain length ($z-$direction) & $1000$\\
$\Delta X$ & grid spacing ($x-$direction)& $20$\\
$\Delta Y$ & grid spacing ($y-$direction)& $20$\\
$\Delta Z$ & grid spacing ($z-$direction)& $20$\\
$\Delta t$ & time step & $180$\\
$\max_{\mathrm{iter}}$ & max. number of iter. for calculation of tissue deformation per time step & $10$\\
\noalign{\smallskip}\hline
\end{tabular}
\end{table}

%
%
\section{Additional Simulation Results}
\begin{figure}[ht!]
\begin{center}
\includegraphics[width=10.5cm]{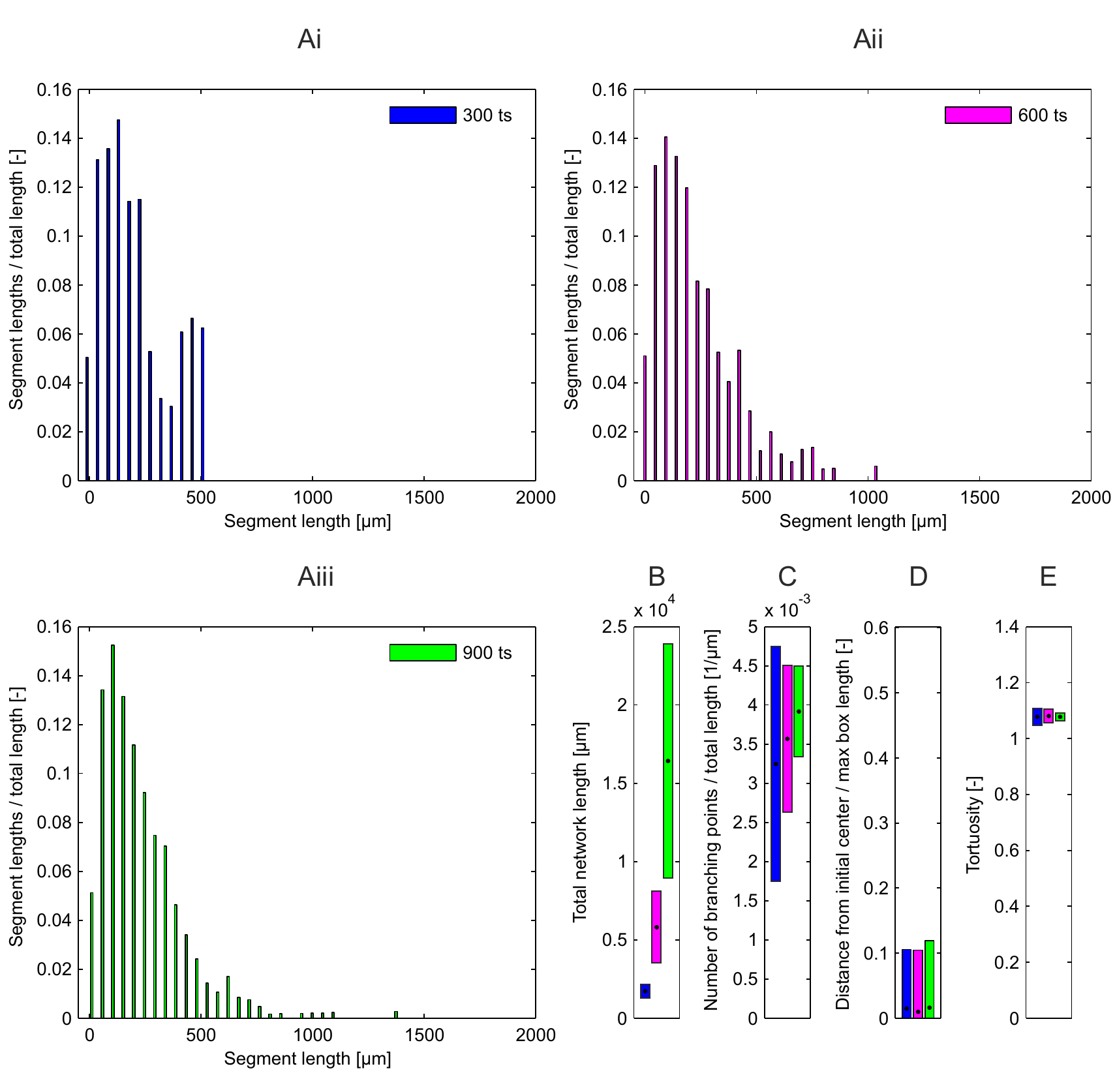}\\
\caption{{\bf Analysis of temporal network evolution.} The metrics stabilise over time, while the total network length grows exponentially.  {\bf(Ai--iii)} the distribution of vessel lengths for $t=300,600,900\,\mathrm{ts}$, {\bf(B)} the total network length, {\bf(C)} the number of branches per unit length; {\bf(D)}, the displacement of the centre of mass, {\bf(E)} the tortuousity. Key: in {\bf (B)--(D)}, means illustrated by dots and standard deviations illustrated by bars were obtained by averaging over $70$ simulations. Parameter values: as per Table~\ref{tab:pars}, except that the chemotactic sensitivity of the tip cells is $\chi =0.0$.}
\label{fig:VesselLengthsTime8}
\end{center}
\end{figure}

\begin{figure}[ht!]
\begin{center}
\includegraphics[width=10.5cm]{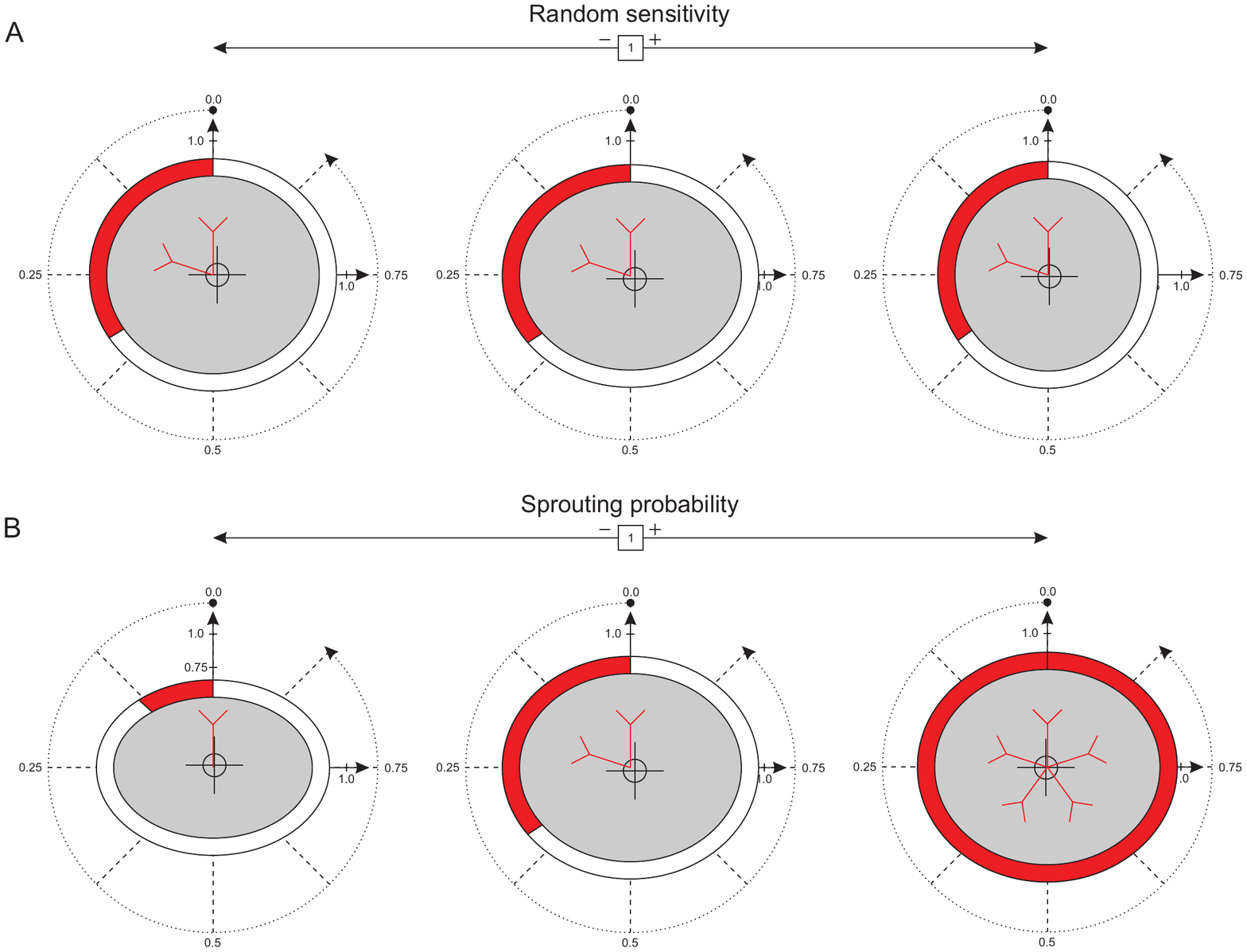}\\
\caption{{\bf Series of glyphs showing how the network metrics at time $t =900\,\mathrm{ts}$ depend on the strength of random sensitivity force and the sprouting probability.} 
Each glyph represents the mean of 70 simulations at time $t =900\,\mathrm{ts}$.
{\bf (A)} the mean behaviour of the networks is does not change as the strength of the random motility force $\sigma$ varies; {\bf (B)} the total network length and the average number of branches per unit vessel length increase as the sprouting probability $k_{\mathrm{spr}}$ increases. For an explanation of the glyphs, see Figure~\ref{fig:GlyphExpl}.
Parameter values: as per Table S.1, except $\chi=0$ and 
(A) $\sigma = 0.2, 0.4, 0.8$ and $k_{\mathrm{spr}} = 2.0 \times 10^{-4}$;
(B) $k_{\mathrm{spr}} = 1.0 \times 10^{-4}, 2.0 \times 10^{-4}, 4 \times 10^{-4}$ and $\sigma = 0.4$. }
\label{fig:Glyphnetwork}
\end{center}
\end{figure}

\begin{figure}[ht]
\begin{center}
\includegraphics[width=11.5cm]{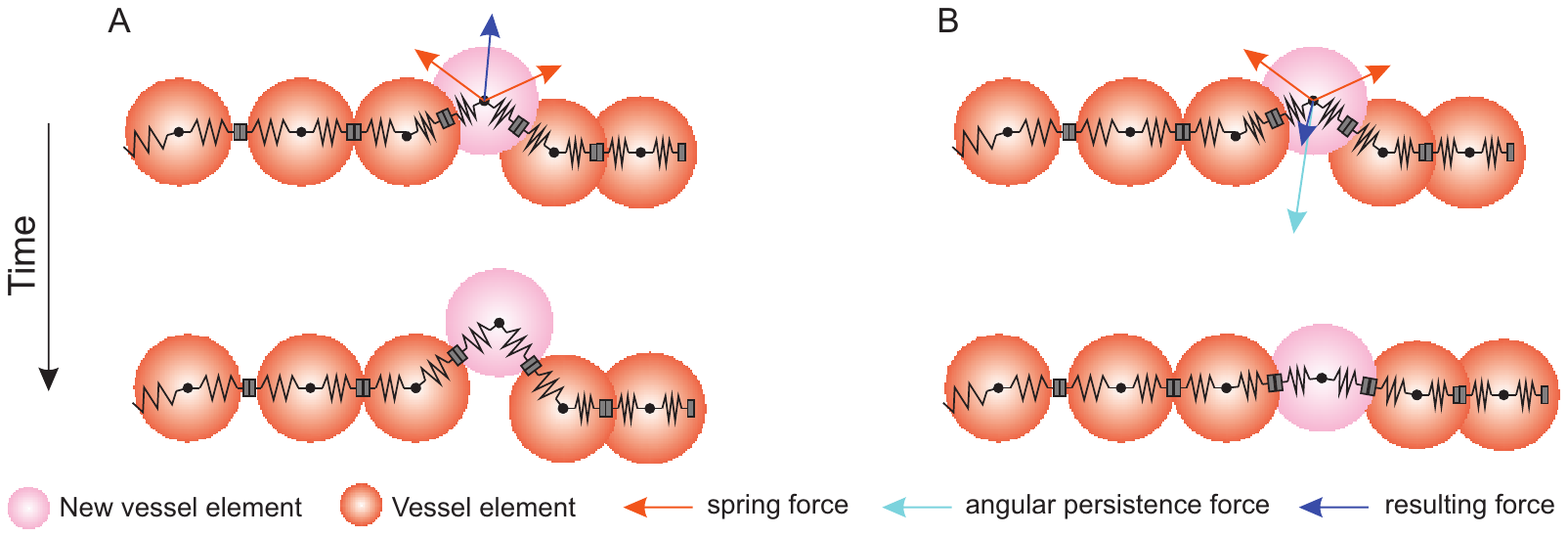}\\
\caption{{\bf The stabilising effect of the angular persistence force on vessel morphology.} This figure illustrates how the angular persistence force stabilises vessel morphology after cell proliferation. {\bf(A)} The new vessel element is pushed out of the vessel if the angular persistence force is neglected. {\bf(B)} when active, the angular persistence force (see Equation~(\ref{angularpersistenceForce})) prevents the new element from being pushed outwards from the parent vessel, leading, instead, to elongation of that vessel.}
\label{fig:AngularPersistence}
\end{center}
\end{figure}

\begin{figure}[ht]
\begin{center}
\includegraphics[width=11.5cm]{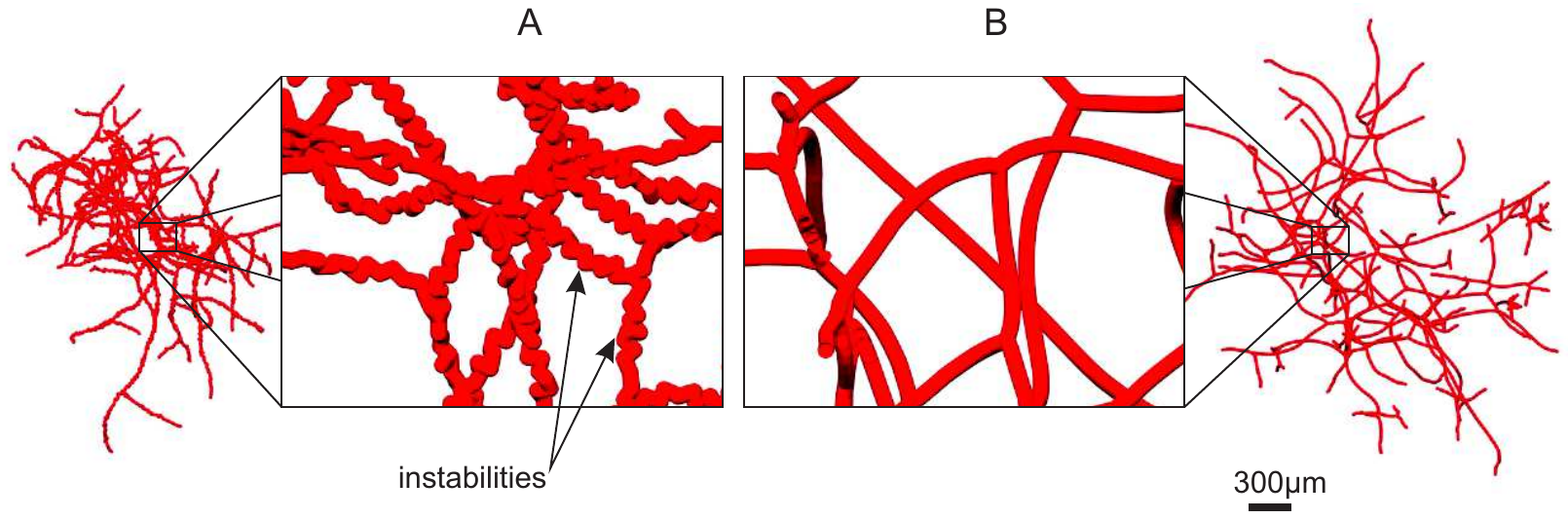}\\
\caption{{\bf Angular persistence forces stabilise vascular networks.} {\bf(A)} If the angular persistence force is inactive ($\lambda_{S_a}=0$), then the network develops small-scale buckling instabilities that resemble the tortuous structures that characterise many solid tumours. {\bf(B)} When the angular persistence force is included, the resulting vascular networks are smooth. Parameter values: as per Table S.1, except $\chi=0.0$. Typical simulation results are plotted at time $t=900\,\mathrm{ts}$.}
\label{fig:NetworkStabilisation}
\end{center}
\end{figure}

\begin{figure}[ht!]
\begin{center}
\includegraphics[width=10.5cm]{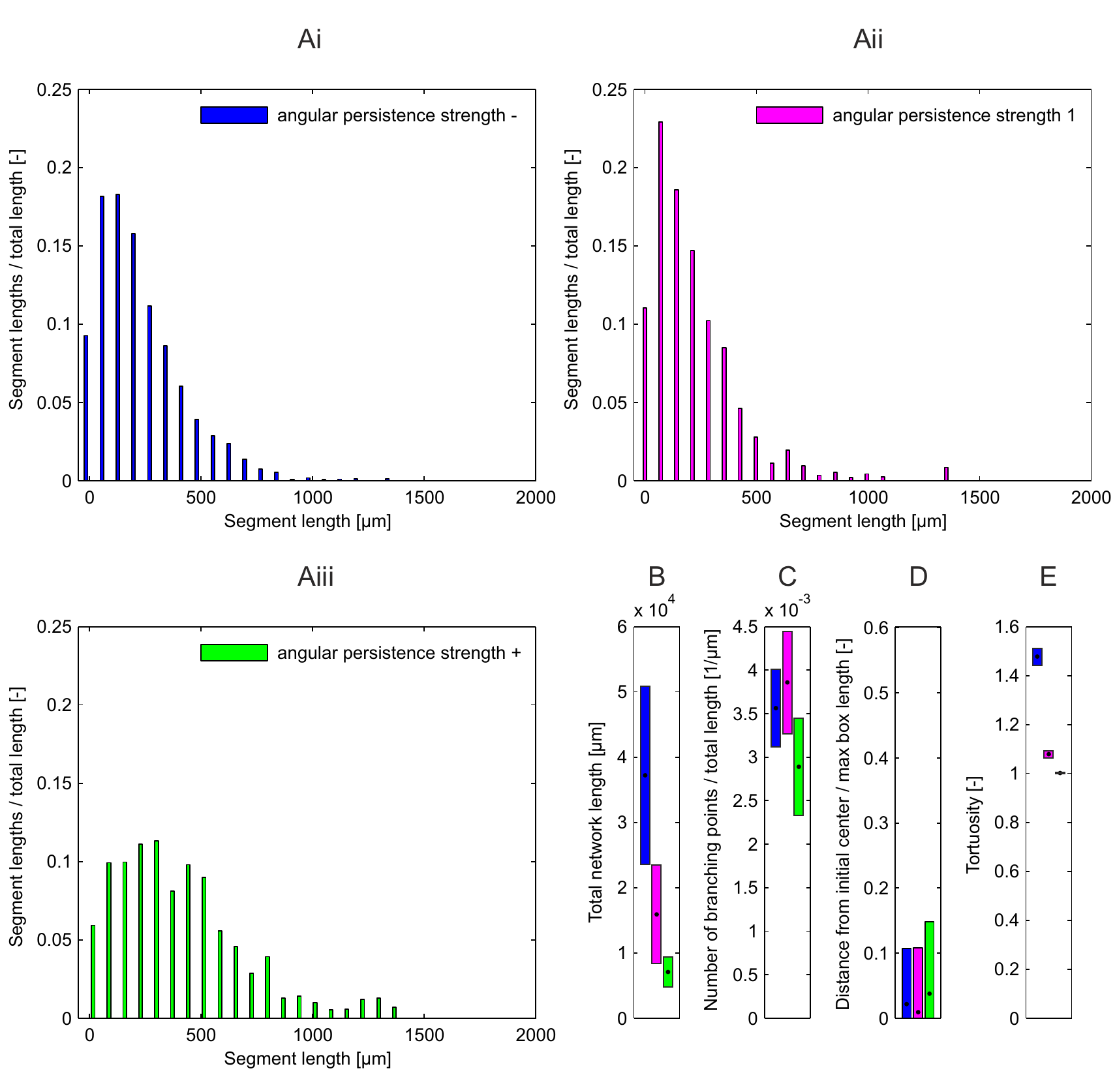}\\
\caption{{\bf Summary statistics showing how the vessel structure at time $t = 900\,\mathrm{ts}$ depends on the strength of the angular persistence force.} 
Parameter sensitivity analysis shows how increasing the strength of the angular persistence force, $\omega_a$, alters the properties of the simulated vascular networks.  
{\bf(Ai--iii)} The distribution of vessel lengths shifts towards longer vessel segments as the angular persistence force increases. {\bf(B, C)} As they became less dense, the total length of the vessel networks and the number of branching points per unit length decrease. 
For each choice of parameter values, summary statistics were obtained by averaging over $70$ simulations. 
Parameter values: as per Table S.1, except $\chi=0$, 
$\omega_a = 5.56 \times 10^{-7}$ (blue),
$\omega_a = 5.56 \times 10^{-5}$ (magenta),
$\omega_a = 5.56 \times 10^{-3}$ (green).}
\label{fig:VesselLengthsTortuosity}
\end{center}
\end{figure}

\begin{figure}[ht!]
\begin{center}
\includegraphics[width=10.5cm]{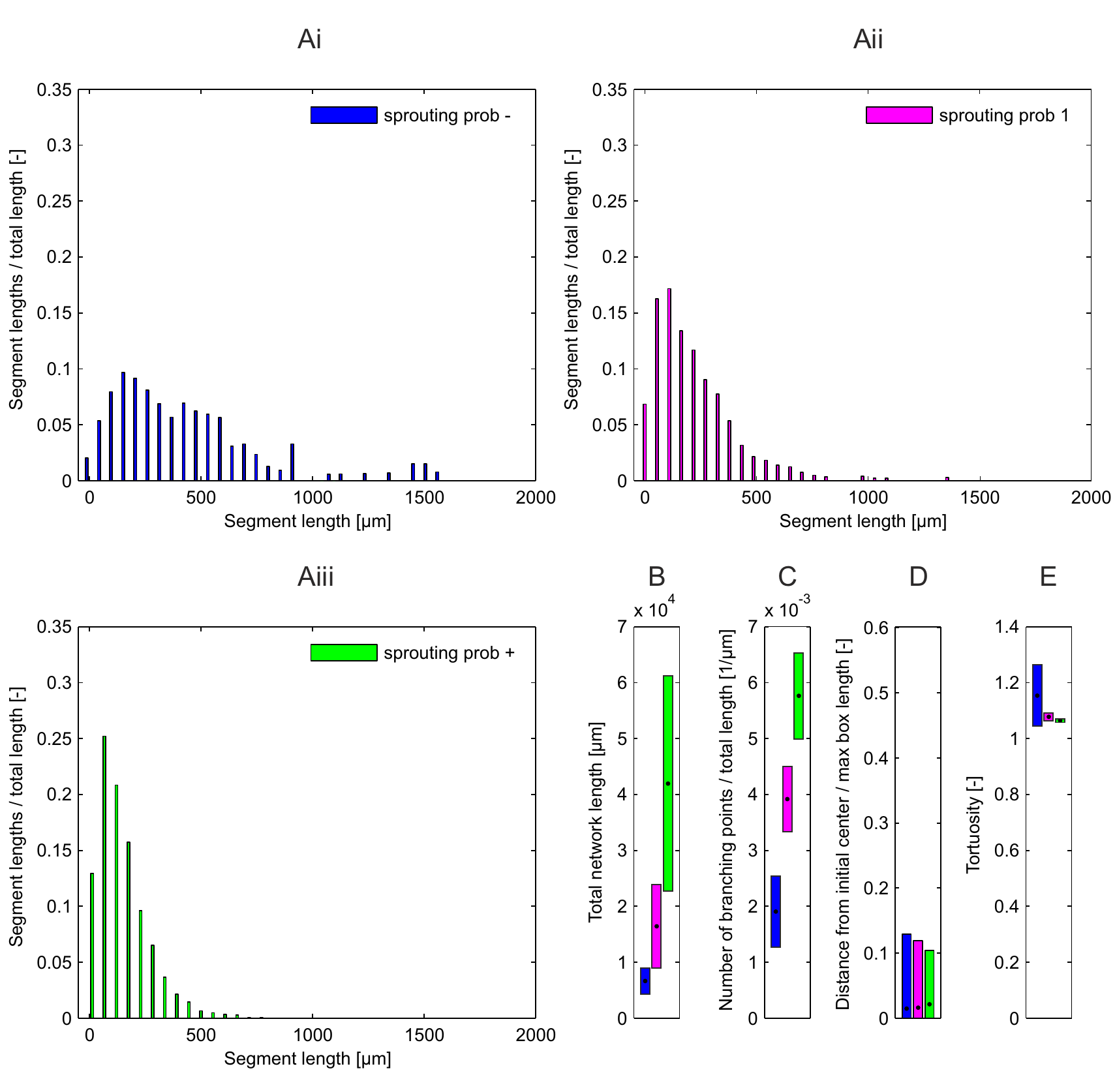}\\
\caption{{\bf Summary statistics showing how network structure at time $t=900\,\mathrm{ts}$ depends on the sprouting probability.}
The figure shows {\bf(Ai--iii)} the distribution of vessel lengths, {\bf(B)} the total network length, {\bf(C)} the number of branches per unit vessel length, {\bf(D)} the displacement of the centre of mass and {\bf(E)} the tortuosity for different sprouting probabilities. 
The metrics which are most sensitive to variation in the sprouting probability
are the total network length {\bf(B)}, and the number of branch points per unit vessel length {\bf(C)}. For each choice of parameter values, summary statistics were obtained by averaging over $70$ simulations. Parameter values: as per Table S.1, except $\chi=0$, $k_{\mathrm{spr}} = 1.0 \times 10^{-4}$ (blue), $k_{\mathrm{spr}} = 2.0 \times 10^{-4}$ (magenta), $k_{\mathrm{spr}} = 3.0 \times 10^{-4}$ (green).}
\label{fig:VesselLengths789}
\end{center}
\end{figure}

In this section, we include additional simulation results \textcolor{black}{Figures \ref{fig:VesselLengthsTime8}--\ref{fig:VesselLengths789}} that complement those presented in the main text.
 
\end{document}